\documentclass[pdflatex,sn-mathphys-num]{sn-jnl}

\usepackage{graphicx}%
\usepackage{multirow}%
\usepackage{amsmath,amssymb,amsfonts}%
\usepackage{amsthm}%
\usepackage{mathrsfs}%
\usepackage[title]{appendix}%
\usepackage{xcolor}%
\usepackage{textcomp}%
\usepackage{manyfoot}%
\usepackage{booktabs}%
\usepackage{algorithm}%
\usepackage{algorithmicx}%
\usepackage{algpseudocode}%
\usepackage{listings}%
\usepackage{greekbf}
\usepackage{todonotes}

\usepackage{hyperref}
\hypersetup{bookmarks=false,
colorlinks=true,
linkcolor=black,
urlcolor=black,
citecolor=black,
pagebackref=true,
pdfstartview={FitH},
bookmarksopenlevel=2
}

\DeclareMathAlphabet{\mathbf}{OT1}{cmr}{bx}{it}
\DeclareMathAlphabet{\bfit}{OT1}{cmr}{bx}{it}

\definecolor{red}{rgb}{0.9,0,0}
\definecolor{blue}{rgb}{0.2,0.2,0.8}
\definecolor{green}{rgb}{0.0,0.5,0.2}
\definecolor{darkblue}{rgb}{0.2,0.2,0.5}
\definecolor{orange}{rgb}{1,0.5,0}

\newcommand{\Orth}{\mathrm{Orth}}
\newcommand{\Skw}{\mathrm{Skw}}

\newcommand{\vers}{\mathop{\textrm{vers}}}

\newcommand {\Real} {\mathbb{R}}
\newcommand {\pb} {\mathbf{p}}
\newcommand {\cb} {\mathbf{c}}
\newcommand {\eb} {\mathbf{e}}
\newcommand {\mb} {\mathbf{m}}
\newcommand {\nb} {\mathbf{n}}
\newcommand {\qb} {\mathbf{q}}
\newcommand {\rb} {\mathbf{r}}
\newcommand {\tb} {\mathbf{t}}
\newcommand {\ub} {\mathbf{u}}
\newcommand {\vb} {\mathbf{v}}
\newcommand {\wb} {\mathbf{w}}
\newcommand {\fb} {\mathbf{f}}
\newcommand {\alphab} {\mathbf{\alpha}}
\newcommand {\deltab} {\mathbf{\delta}}
\newcommand {\lambdab} {\mathbf{\lambda}}
\newcommand {\omegab} {\mathbf{\omega}}
\newcommand {\Ab} {{\bf A}}
\newcommand {\Bb} {\mathbf{B}}
\newcommand {\Cb} {\mathbf{C}}
\newcommand {\Ib} {\mathbf{I}}
\newcommand {\Rb} {{\bf R}}
\newcommand {\Wb} {{\bf W}}
\newcommand {\0} {{\bf 0}}
\newcommand {\Bc} {\mathcal{B}}

\newcommand{\eef}{\mathfrak{e}}
\newcommand{\gef}{\mathfrak{g}}

\newcommand{\bigo}[1]{\setbox0\hbox{$\bigcirc$}%
             \rlap{\raise .2ex\hbox to \wd0{\hfil ${\scriptscriptstyle
                   #1}$\hfil}}\box0}

\graphicspath{{./}{Figures/}}

\begin{document}


\title{Simulation and design of isostatic thick 
origami structures}


\author*[1]{\fnm{Andrea } \sur{Micheletti}}\email{micheletti@ing.uniroma2.it}

\author[1]{\fnm{Alessandro} \sur{Tiero}}\email{tiero@uniroma2.it}

\author[2]{\fnm{Giuseppe} \sur{Tomassetti}}\email{giuseppe.tomassetti@uniroma3.it}

%

\affil*[1]{\orgdiv{Dipartimento di Ingegneria Civile e Ingegneria Informatica},\\ \orgname{University of Rome Tor Vergata}, \orgaddress{\street{Via Politecnico 1}, \city{Rome}, \postcode{00133}, 
\country{Italy}}}

\affil[2]{\orgdiv{Dipartimento di Ingegneria Industriale, Elettronica e Meccanica},\\ \orgname{Roma Tre University}, \orgaddress{\street{Street}, \city{Rome}, \postcode{00100}, 
\country{Italy}}}




\abstract{
Thick origami structures are considered here as assemblies of polygonal panels hinged to each other along their edges according to a corresponding origami crease pattern. The determination of the internal actions caused by external loads in such structures is not an easy task, owing to their high degree of static indeterminacy, and the likelihood of unwanted  self-balanced internal actions induced by manufacturing imperfections. 
Here we present a method for reducing the degree of static indeterminacy which can be applied to several thick origami structures to make them isostatic. The method utilizes sliding hinges, which permit also the relative translation along the hinge axis, to replace conventional hinges.
After giving the analytical description of both types of hinges 
and describing a rigid folding simulation procedure based on the integration of the exponential map, we present the static analysis of a series of noteworthy examples based on the Miura-ori pattern, the Yoshimura pattern, and the Kresling pattern.
The method can be applied for the design and realization of thick origami structures with adequate strength to resist external actions.


}




\maketitle

\section{Introduction}



Structures realized as or inspired by origami interested several researchers in different fields~\cite{Meloni2021}. For example, architectural applications were presented in~\cite{Ljubas2010,Reis2015,DAcunto2018}, while metamaterial applications were described in~\cite{Evans2015,Brunck2016,Overvelde2016}; light-activated origami folding was proposed in~\cite{Liu2011,Lee2015,Liu2017}, and actuation by uniform heating was achieved in~\cite{Tolley2014}.

The physical realization of transformable origami-like structures requires to obtain reliable predictions regarding the folding/deploying process and the stiffness and strength under 
external loads.
Numerous studies addressed these tasks for thin origami structures~\cite{Lebee2015}, i.e., those obtained by folding a thin continuum layer of material: simulation procedures for rigid and non-rigid origami folding were proposed by several authors \cite{Lang1996,Miyazaki1996,Demaine2007,Tachi2010,Xi2015}, while mechanical models adopting various Stick-and-Spring idealizations \cite{Favata2014} were proposed in \cite{Schenk2011,Tachi2013,Schenk2013,Evans2015,Magliozzi2017,Liu2017a,Ghassaei2018}.

In contrast to thin origami structures, thick origami structures are 
assembled from a certain number of polygonal panels, hinge-connected to each other along their edges, according to a given crease pattern.
Thick origami structures are typically overconstrained, that is, they are statically underdetermined structures, or, in other words, they possess several self-stress states. This poses the problem of the likelihood of unwanted self-balanced internal actions induced by manufacturing imperfections.
At the same time, the choice of constitutive relations for the internal actions associated with well defined strain measures is not trivial. For these reasons, to perform the structural design of these systems is a challenging task. 

In this work, we detail a strategy for reducing the degree of static indeterminacy of thick origami structures, possibly bringing it down to zero, in order to make them isostatic structures. We take advantage of the introduction of sliding hinges, in addition to conventional door hinges, in a rigid origami model.
Sliding hinges provide an additional degree of freedom with respect to door hinges, in that they permit the relative translation between adjacent panels along the shared hinge axis, other than the relative rotation between the two panels about the same axis.
Therefore, by replacing a door hinge by a sliding hinge in a thick origami realization, either one degree of static indeterminacy is removed or one degree of kinematic indeterminacy is added.
We found several noteworty cases in which the present strategy is successful in making the thick origami structure isostatic, so that the internal actions induced by external loads can be uniquely computed.

Preliminary result of our procedure were presented in \cite{Micheletti2022} for the case of a Yoshimura crease pattern.
Here, other than providing the details of the modeling equations and the simulation algorithm, we present results concerning the statics of thick Miura-ori, Yoshimura, and Kresling origami.

In the following, after recalling useful counting rules for the degrees of kinemaic and static indeterminacy (Section 2),
we give an exact finite kinematic description of the motion of rigid origami structures with door hinges and sliding hinges, together with the infinitesimal counterpart (Sections 3.1\,-\,3.3).
Folding/deployment simulations can then be performed by numerical integration of the exponential map, by the procedure outlined in Section 3.4.
The equilibrium equations relating external loads and internal actions are obtained by duality 
(Section 4.1). Afterward, the equilibrium equations are solved numerically for selected examples to demonstrate the usefulness of the method (Section 4.2).
We close by discussing the obtained results and future extensions (Section 5).

\section{Counting mechanisms and self-stress states}

In this section, some handy results on the number of internal mechanisms and self-stress states of rigid origami structures, which corresponds respectively to the degrees of kinematic and static indeterminacy, are reviewed for later use (cf.~\cite{Micheletti2022}).

A starting point is provided by the {\em convex triangulated polyhedron theorem}~\cite{Whiteley1999}.
The theorem states that a pin-jointed bar framework constructed on a convex triangulated polyhedron, by placing the bars and pins of the framework on the edges and vertices of the polyhedra, is isostatic. Then, by removing one edge from such a framework, an internal mechanism is produced, while the four edges surrounding the removed edge form the boundary of a so-called {\em hole}~\cite{FinbowSingh2012}. Next, by removing one of those four edges, a bigger five-edge hole is formed and another internal mechanism is introduced. Hence, the number of mechanisms introduced by a hole bounded by $E_b$ edges is $E_b-3$. By observing that a origami with a triangulated crease pattern has the same edge-vertex connectivity of a convex triangulated polyhedron possessing one hole with a number of boundary edges equals to the number of boundary edges of the origami, one can conclude that a pin-jointed framework constructed on a triangulated origami crease pattern, in a non-singular configuration, has $E_b-3$ independent internal mechanism.

In order to determine the number of self-stress states, one can consider again a framework built on a convex triangulated polyhedron and add one edge between the two distant vertices of two adjacent triangles. In this case, the four vertices of the two adjacent triangles form a so-called {\em block}, and the framework acquires one self-stress state. By matching the polygonal panels of the origami with the blocks of the pin-connected framework, it is easy to see that each polygonal panel of the origami with $V$ vertices corresponds to $V-3$ independent self-stress states in the framework.

\begin{figure}[h!]
\centering
\includegraphics[width=0.9\textwidth]{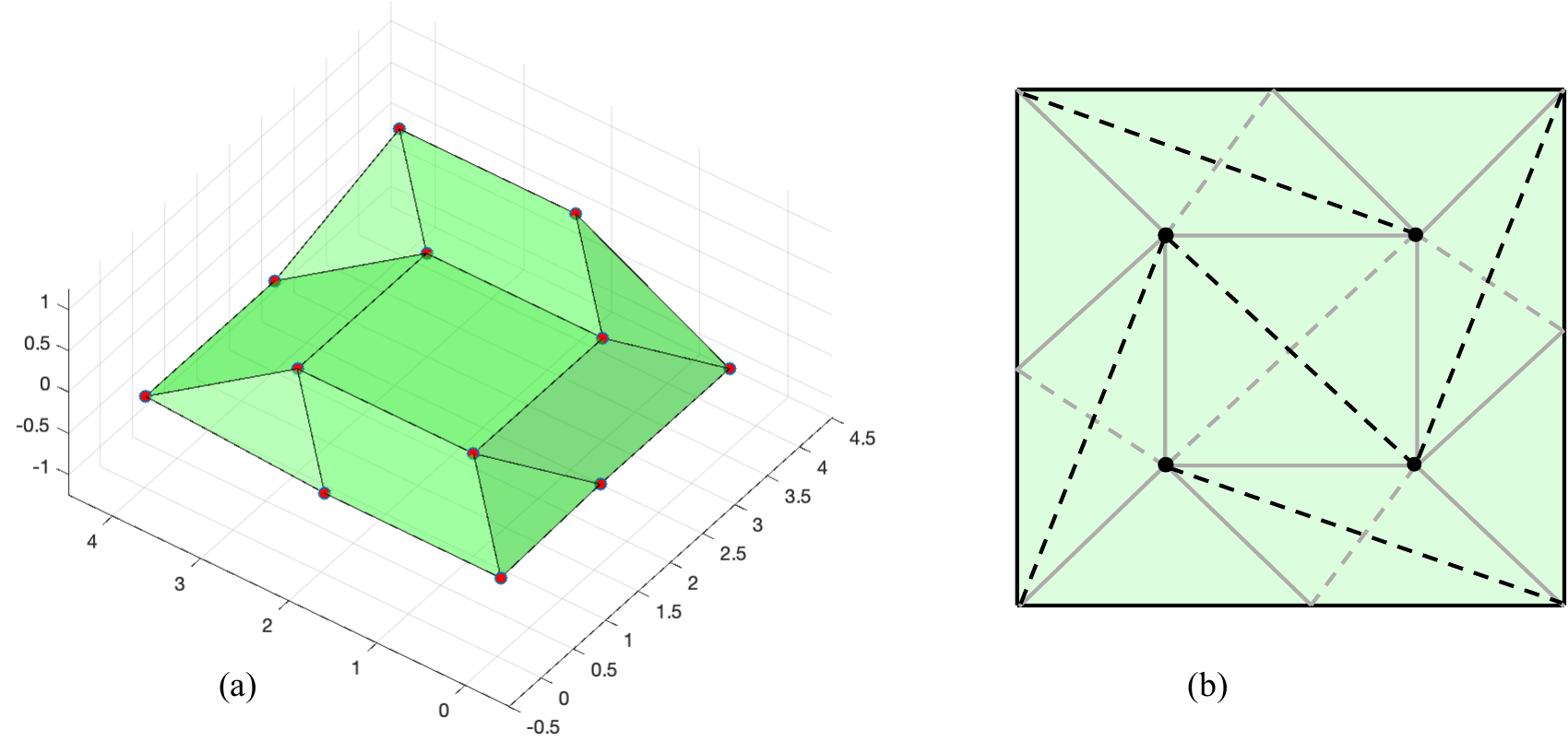}
\caption{The square twist origami (a) and the corresponding bar framework idealization (b). There are $P=9$ panels, $H=12$ hinge edges (solid light gray lines), and  $E_b=8$ boundary edges (solid black lines). Each of the five quadrilateral panels is triangulated with one additional edge (dashed light gray line) and transformed into a block by adding one more edge (dashed black line). There are $V_i=4$ internal vertices (black dots). The bar framework model of this origami has $S_{BF}=1$ self-stress state, and the panel-hinge model has $S_{PH}=13$, while $M=1$ in both models. 
}\label{bh}
\end{figure}

By considering the kinematics of a panel-hinge model, in which only the relative rotation about the common edge between adjacent panels is permitted, one can arrive at the same conclusions of the bar framework model. However, the predictions regarding the statics differ in the two models. In fact, it is easy to see that the panel-hinge model gives a higher number of self-stress states. In particular, it was shown in~\cite{Micheletti2022} that $S_{PH}\,=\,S_{BF}\,+\,3\,V_i$, with $S_{PH}$ and $S_{BF}$ the degree of static indeterminacy computed according to the panel-hinge and bar-framework models, respectively, and $V_i$ the number of the internal vertices of the origami. 
Figure~\ref{bh} illustrates these quantities for the square twist origami.

\section{Kinematics}\label{sec:kin}

In this section, we first derive the exact constraint equations for sliding hinges and door hinges between adjacent panel-shaped bodies, together with the corresponding linearized versions. Then we give them an equivalent description by composition of point slider and spherical hinge constraints. After that we present an integration procedure of the kinematic equations based on the Newton-Raphson algorithm.
In the following, we consider the motion of a body $\Bc$ in the Euclidean space with respect to a fixed reference frame as described by the rigid motion $g=(\Rb,\ub)$ in terms of a rotation tensor $\Rb\in\Orth^+$ and a translation vector $\ub\in V$.

Let $\Bc_1$ and $\Bc_2$ be two bodies connected to each other by a hinge with axis $r$ in the reference configuration, and let $r_1$ and $r_2$ the images of $r$ in the rigid motions $g_1=(\Rb_1,\ub_1)$ and $g_2=(\Rb_2,\ub_2)$ of $\Bc_1$ and $\Bc_2$ respectively:
\begin{align*}
\label{rigmot1}
& r_1=\{\Rb_1(\pb+s_1 \tb)+\ub_1, \, s_1\in\Real\}\,,\\
& r_2=\{\Rb_2(\pb+s_2 \tb)+\ub_2 ,\, s_2\in\Real\}\,,
\end{align*}
with $\pb$ is the position vector of a point $P$ on $r$, and $\tb$ is the unit vector parallel to $r$.

\subsection{Sliding hinges}

\

In case of a sliding hinge, we have that $r_1=r_2$, that is,
$\forall s_1\in\Real, \exists s_2\in\Real$, or
$\forall s_2\in\Real, \exists s_1\in\Real$, such that  
\begin{equation}
\Rb_1(\pb+s_1 \tb)+\ub_1=\Rb_2(\pb+s_2 \tb)+\ub_2\,,
\label{slidingHingeConstraint}
\end{equation}
that is
$$
s_2(s_1) \Rb_2 \tb - s_1 \Rb_1 \tb + \Rb_2 \pb + \ub_2 - \Rb_1 \pb - \ub_1 = \0\,,
$$
or 
$$
s_2 \Rb_2 \tb - s_1(s_2) \Rb_1 \tb + \Rb_2 \pb + \ub_2 - \Rb_1 \pb - \ub_1 = \0\,,
$$
with $s_1 \mapsto s_2(s_1)$ and $s_2 \mapsto s_1(s_2)$ and $s_1 \circ s_2$ the identity. 
By differentiating with respect to the parameter $s_1$ the first of these two equivalent relations, we get
$
s_2^\prime  \Rb_2 \tb -  \Rb_1 \tb = \0 \,,
$
so that  $ {s_2}^\prime = \pm 1$ and  $\Rb_2 \tb = \pm  \Rb_1 \tb$.
Assuming that at time $t_0$,  $\Rb_2(t_0) \tb =  \Rb_1(t_0) \tb$,
the continuity of the motion implies that
$$
                   \Rb_2 \tb =  \Rb_1 \tb =  \alphab \,,
$$
with $\alphab$ an arbitrary vector.
Then, substitution in (\ref{slidingHingeConstraint}) yields
$$
 \alphab \ \times \  (\pb_2-\pb_1)  = \0 \,,
$$
where  $\pb_1$ and $\pb_2$ denote the image of the position vector $\pb$ under the rigid motios $g_1$ and $g_2$ of the two bodies:
$$\pb_1=\Rb_1 \pb + \ub_1\,,\quad \pb_2=\Rb_2 \pb + \ub_2\,.$$
%
%
The constraints (\ref{slidingHingeConstraint}) can be realized, for example, by the scalar equations
\begin{equation}
\Rb_2 \tb \cdot \Rb_1 \mb =0\,,\>
\Rb_2 \tb \cdot \Rb_1 \nb =0\,,\>
(\pb_2-\pb_1) \cdot \Rb_1 \mb=0\,,\>
(\pb_2-\pb_1) \cdot \Rb_1 \nb=0\,,  \nonumber
\end{equation}
or, by the scalar equations,
\begin{equation}
\Rb_1 \tb \cdot \Rb_2 \mb =0\,,\>
\Rb_1 \tb \cdot \Rb_2 \nb =0\,,\>
(\pb_2-\pb_1) \cdot \Rb_2 \mb=0\,,\>
(\pb_2-\pb_1) \cdot \Rb_2 \nb=0\,, \nonumber
\end{equation}
where $\nb, \mb$ are  two  linearly independent unit vectors orthogonal to $\tb$.

We choose as constraint equations  the   arithmetic mean of equations of the two groups. Consequently,  
the constraint functions are 
\begin{eqnarray}
c_1 &=&  \frac{1}{2}( \tb_1 \cdot  \mb_2  +   \tb_2 \cdot  \mb_1  ) \,, \nonumber \\
c_2  &=& \frac{1}{2}(   \tb_1 \cdot  \nb_2 +    \tb_1  \cdot  \nb_2 )  \,, \nonumber \\
c_3  &=&\frac{1}{2}(  (\pb_2-\pb_1) \cdot \mb_1 + (\pb_2-\pb_1) \cdot  \mb_2) \,, \nonumber \\
c_4  &=&\frac{1}{2}(  (\pb_2-\pb_1) \cdot  \nb_1 + (\pb_2-\pb_1) \cdot  \nb_2)\,,\>\nonumber 
\end{eqnarray}
with  $\mb_i = \Rb_i \mb$,  $\nb_i =  \nb$, $\tb_i = \Rb_i \tb$, $i=1,2$. 
%
%
The time derivatives of the constraints functions have the expressions 
\begin{eqnarray}
\dot c_1  &=&   \Wb_1 \tb_1 \cdot  \mb_2  + \tb_1 \cdot \Wb_2 \mb_2 +
                       \Wb_2 \tb_2 \cdot  \mb_1  + \tb_2 \cdot  \Wb_1 \mb_1    \nonumber \\
             &=&    (\Wb_1 - \Wb_2) \tb_1 \cdot  \mb_2  +
                       (\Wb_2 - \Wb_1) \tb_2 \cdot  \mb_1  \,,    \nonumber \\ [5pt]
\dot c_2  &=&   \Wb_1 \tb_1 \cdot  \nb_2  + \tb_1 \cdot \Wb_2 \nb_2 +
                      \Wb_2 \tb_2 \cdot  \nb_1  + \tb_2 \cdot  \Wb_1 \nb_1    \nonumber \\
             &=&    (\Wb_1 - \Wb_2) \tb_1 \cdot  \nb_2  +
                       (\Wb_2 - \Wb_1) \tb_2 \cdot  \nb_1  \,,    \nonumber \\ [5pt]
\dot c_3  &=& (\Wb_2 \pb_2 - \Wb_1\pb_1 + \wb_2 -\wb_1) \cdot \mb_1 
                     +   (\pb_2-\pb_1 ) \cdot \Wb_1 \mb_1   +                                  \nonumber   \\
                    &+& (\Wb_2 \pb_2 - \Wb_1\pb_1 + \wb_2 -\wb_1) \cdot \mb_2 
                    + (\pb_2-\pb_1 ) \cdot \Wb_2 \mb_2 \,, \nonumber                        \nonumber \\
            &=& (\Wb_2 -\Wb_1) \pb_2  \cdot \mb_1  + (\Wb_2 - \Wb_1) \pb_1 \cdot \mb_2 
                   + (\wb_2 -\wb_1) \cdot( \mb_1+ \mb_2) \,, \nonumber      \\   [5pt]          
\dot c_4  &=& (\Wb_2 \pb_2 - \Wb_1\pb_1 + \wb_2 -\wb_1) \cdot \nb_1 
                     +   (\pb_2-\pb_1 ) \cdot \Wb_1 \nb_1   +                                  \nonumber   \\
                    &+& (\Wb_2 \pb_2 - \Wb_1\pb_1 + \wb_2 -\wb_1) \cdot \nb_2 
                    + (\pb_2-\pb_1 ) \cdot \Wb_2 \nb_2 \,, \nonumber                        \nonumber \\
            &=& (\Wb_2 -\Wb_1) \pb_2  \cdot \nb_1  + (\Wb_2 - \Wb_1) \pb_1 \cdot \nb_2 
                   + (\wb_2 -\wb_1) \cdot( \nb_1+ \nb_2) \,. 
\label{slidingHingeDerivatives}
\end{eqnarray}
To obtain these expressions we have used the differentiation rules
\begin{equation}
  \dot\pb_i = \Wb_i \pb_i+\wb_i \,,\>  \dot\tb_i = \Wb_i \tb_i \,, \> 
\dot\mb_i = \Wb_i \mb_i \,,\> \dot\nb_i = \Wb_i \nb_i  \,,\quad i=1,2\,, \nonumber
\end{equation}
with $\Wb_i = \dot \Rb_i \Rb_i^T$ and $\wb_i = \dot \ub_i - \Wb_i \ub_i$.
By introducing the angular velocities $\omegab_i$ as the axial vectors of $\Wb_i$, $i=1,2$,  (\ref{slidingHingeDerivatives}) are expressed as follows
\begin{eqnarray}
\dot c_1   &=&  ( \tb_1 \times  \mb_2 + \tb_2 \times  \mb_1)   \cdot ( \omegab_2 - \omegab_1) \,,    \nonumber \\
\dot c_2    &=&  (\tb_1 \times \nb_2  + \tb_2  \times  \nb_1)   \cdot ( \omegab_2 - \omegab_1)  \,,    \nonumber \\
\dot c_3   &=&  ( \pb_2  \times \mb_1   + \pb_1 \times \mb_2) \times( \omegab_2 - \omegab_1)
                   + (\wb_2 -\wb_1) \cdot (\mb_1+ \mb_2) \,, \nonumber      \\            
\dot c_4  &=&  ( \pb_2  \times \nb_1   + \pb_1 \times \nb_2) \times( \omegab_2 - \omegab_1)
                   + (\wb_2 -\wb_1) \cdot ( \nb_1+ \nb_2) \,. \nonumber        
\end{eqnarray}

Then, the first order approximation of the  constraint system  in 
 the neighborhood of the configuration $(\Rb_1 \,, \ub_1\,,\Rb_2 \,, \ub_2)$,  has the expression
\begin{equation}
                              \cb_0 + \Cb \deltab = \0 \,,
\end{equation}
with $\cb = (c_1\,, c_2\,, c_3\,, c_4)$,   $\cb_0 = \cb(\Rb_1 \,, \ub_1\,,\Rb_2 \,, \ub_2)$, 
$\deltab = (\delta \omegab_1\,, \delta \wb_1\,, \delta \omegab_2\,, \delta \wb_2)$,
and
\begin{equation}\label{slid:C}
\Cb = \frac{1}{2}
\left(
\begin{array}{cccc}
 \scriptstyle -(\tb_1 \times  \mb_2 + \tb_2 \times  \mb_1)^T  & \scriptstyle \0  &
     \scriptstyle ( \tb_1 \times  \mb_2 + \tb_2 \times  \mb_1)^T  &  \scriptstyle \0 \\ 
 \scriptstyle -(\tb_1 \times \nb_2  + \tb_2  \times  \nb_1)^T  &  \scriptstyle \0  &
     \scriptstyle (\tb_1 \times \nb_2  + \tb_2  \times  \nb_1)^T    &  \scriptstyle  \0 \\
\scriptstyle- ( \pb_2  \times \mb_1   + \pb_1 \times \mb_2)^T & \scriptstyle -(\mb_1+ \mb_2)^T &
     \scriptstyle ( \pb_2  \times \mb_1   + \pb_1 \times \mb_2)^T & \scriptstyle (\mb_1+ \mb_2)^T  \\
\scriptstyle -( \pb_2  \times \nb_1   + \pb_1 \times \nb_2)^T & \scriptstyle -( \nb_1+ \nb_2)^T &
    \scriptstyle  ( \pb_2  \times \nb_1   + \pb_1 \times \nb_2)^T & \scriptstyle ( \nb_1+ \nb_2)^T 
\end{array}
\right)\,. 
\end{equation}

\subsection{Door hinges}

For door hinges we start with the condition
$$
\forall s\in\Real\,, \quad \Rb_1(\pb+s \tb)+\ub_1=\Rb_2(\pb+s \tb)+\ub_2\,.
$$
With the notation used in the previous section,
\begin{align*}
& \tb_2 - \tb_1 =\0\quad\mathrm{(as\ in\ the\ case\ of\ a\ sliding\ hinge)}\,,\\
& \pb_2 - \pb_1=\0\,.
\end{align*}
Then the constraint function are
\begin{eqnarray}
c_1 &=& ( \tb_1 \cdot  \mb_2  +   \tb_2 \cdot  \mb_1  ) \,, \nonumber \\
c_2  &=& (   \tb_1 \cdot  \nb_2 +    \tb_1  \cdot  \nb_2 )  \,, \nonumber \\
\cb_3  &=&   \pb_2-\pb_1 \,. \nonumber
\end{eqnarray}
The time derivatives of the constraints functions have the expressions 
\begin{eqnarray}
\dot c_1  &=&   \Wb_1 \tb_1 \cdot  \mb_2  + \tb_1 \cdot \Wb_2 \mb_2 +
                       \Wb_2 \tb_2 \cdot  \mb_1  + \tb_2 \cdot  \Wb_1 \mb_1    \nonumber \\
             &=&    (\Wb_1 - \Wb_2) \tb_1 \cdot  \mb_2  +
                       (\Wb_2 - \Wb_1) \tb_2 \cdot  \mb_1  \,,    \nonumber \\ [5pt]
\dot c_2  &=&   \Wb_1 \tb_1 \cdot  \nb_2  + \tb_1 \cdot \Wb_2 \nb_2 +
                      \Wb_2 \tb_2 \cdot  \nb_1  + \tb_2 \cdot  \Wb_1 \nb_1    \nonumber \\
             &=&    (\Wb_1 - \Wb_2) \tb_1 \cdot  \nb_2  +
                       (\Wb_2 - \Wb_1) \tb_2 \cdot  \nb_1  \,,    \nonumber \\ [5pt]
\dot \cb_3  &=& (\Wb_2 \pb_2 + \wb_2 - \Wb_1\pb_1 - \wb_1) \,, \nonumber      
\label{doorHingeDerivatives}
\end{eqnarray}
or, in terms of angular velocities, 
\begin{eqnarray}
\dot c_1   &=&  ( \tb_1 \times  \mb_2 + \tb_2 \times  \mb_1)   \cdot ( \omegab_2 - \omegab_1) \,,    \nonumber \\
\dot c_2    &=&  (\tb_1 \times \nb_2  + \tb_2  \times  \nb_1)   \cdot ( \omegab_2 - \omegab_1)  \,,    \nonumber \\
\dot \cb_3   &=&  \omegab_2 \times \pb_2  + \wb_2 -  \omegab_1 \times \pb_2 - \wb_1\,. \nonumber          
\end{eqnarray}
Then, the first order approximation of the  the constraint system  in 
 the neighborhood of the configuration $(\Rb_1 \,, \ub_1\,,\Rb_2 \,, \ub_2)$,  has the expression
\begin{equation}
                              \cb_0 + \Cb \deltab = \0 \,,
\end{equation}
with $\cb = (c_1\,, c_2\,, \cb_3)$,   $\cb_0 = \cb(\Rb_1 \,, \ub_1\,,\Rb_2 \,, \ub_2)$, 
$\deltab = (\delta \omegab_1\,, \delta \wb_1\,, \delta \omegab_2\,, \delta \wb_2)$,
and
\begin{equation}\label{door:C}
\Cb = \frac{1}{2}
\left(
\begin{array}{cccc}
 \scriptstyle -(\tb_1 \times  \mb_2 + \tb_2 \times  \mb_1)^T  & \scriptstyle \0  &
     \scriptstyle ( \tb_1 \times  \mb_2 + \tb_2 \times  \mb_1)^T  &  \scriptstyle \0 \\ 
 \scriptstyle -(\tb_1 \times \nb_2  + \tb_2  \times  \nb_1)^T  &  \scriptstyle \0  &
     \scriptstyle (\tb_1 \times \nb_2  + \tb_2  \times  \nb_1)^T    &  \scriptstyle  \0  \\  
\scriptstyle  *\pb_1 &  \scriptstyle -\Ib
& \scriptstyle -\,*\pb_2 &  \scriptstyle \Ib
\end{array}
\right)\,,
\end{equation}
with $*\pb_i$ the axial tensors of $\pb_i$, $i=1,2$, and $\Ib$ the identity tensor.


\subsection{Point constraints}

We express here the hinge constraint between two bodies in terms of point constraints, that is, a sliding hinge is obtained with two sliders located on the hinge axis, a door hinge is obtained as a fixed point and a slider located on the hinge axis.

\subsubsection{Single point slider}
Let the axis of the guide  be described in the reference configuration by the line $r$ solidal to $\Bc_1$ that passes through the points
$O +\pb$ and $O +\qb$; and 
let  $O + \qb$ be  the point of  $\Bc_2$ constrained to slide on  the guide.
Under the motion of the two bodies  the vector positions $\pb$, $\qb$ are transported to the vectors
\begin{align*}
& \pb_i =  \Rb_i\pb+\ub_i\,, \\
& \qb_i=\Rb_i\qb+\ub_i \,,
\end{align*} 
and the line $r$ to the lines  
\begin{equation} r_i= \{ O+  \Rb_i(\pb+s\tb)+\ub_i\,,\ s\in\Real \} \,,
\end{equation}
with $\tb =  \vers (\qb-\pb)$ and $i=1,2$.

%
%
The constraint imposes that the point $\pb_2$ be the position vector of a point of $r_1$, that is 
$$
\pb_2 - \pb_1 \quad  || \quad  \ r_1\,.
$$
It follows that the constraint equations are as follows:
\begin{eqnarray}
c_1(\Rb_1\,, \ub_1\,, \Rb_2\,, \ub_2)  &=&  (\pb_2 - \pb_1)\cdot\mb_1  \,,   \nonumber\\
c_2(\Rb_1\,, \ub_1\,, \Rb_2\,, \ub_2)  &=&  (\pb_2 - \pb_1)\cdot\nb_1  \,,    \nonumber
\end{eqnarray}
with $\mb_1 = \Rb_1 \mb$ and  $\nb_1 = \Rb_1 \nb$.

%
%
The derivatives with respect to time of these  functions have the expressions 
\begin{eqnarray}
\dot c_1 &=& (\Wb_2  \pb_2 + \wb_2  -  \Wb_1  \pb_1- \wb_1) \cdot \mb_1 +  (\pb_2  -  \pb_1) \cdot \Wb_1 \mb_1 \nonumber \\       
            &=&  \left( (\Wb_2  - \Wb_1) \pb_2 +   \wb_2 - \wb_1 \right) \cdot \mb_1     \nonumber \\   
            &=&  \left( (\Wb_2  - \Wb_1) \pb_2 +   \wb_2 - \wb_1 \right) \cdot \mb_1     \nonumber \\   
            &=&    \pb_2 \times \mb_1 \cdot  (\omegab_2 - \omegab_1) +  \mb_1 \cdot (\wb_2 - \wb_1)\,,   \nonumber \\
\dot c_2 &=& (\Wb_2  \pb_2 + \wb_2  -  \Wb_1  \pb_1- \wb_1) \cdot \nb_1 +  (\pb_2  -  \pb_1) \cdot \Wb_1 \nb_1 \nonumber \\       
            &=&  \left( (\Wb_2  - \Wb_1) \pb_2 +   \wb_2 - \wb_1 \right) \cdot \nb_1     \nonumber \\   
            &=&  \left( (\Wb_2  - \Wb_1) \pb_2 +   \wb_2 - \wb_1 \right) \cdot \nb_1     \nonumber \\   
            &=&    \pb_2 \times \nb_1 \cdot  (\omegab_2 - \omegab_1) +  \nb_1 \cdot (\wb_2 - \wb_1)\,.   \nonumber 
\end{eqnarray}
%
%
Then, the first order approximation of the  constraint equations  in 
 the neighborhood of the configuration $(\Rb_1 \,, \ub_1\,,\Rb_2 \,, \ub_2)$,  has the expression
\begin{equation}
                              \cb_0 + \Cb \deltab = 0 \,,
\end{equation}
with $\cb = (c_1\,, c_2)$,   $\cb_0 = \cb(\Rb_1 \,, \ub_1\,,\Rb_2 \,, \ub_2)$, 
$\deltab = (\delta \omegab_1\,, \delta \wb_1\,, \delta \omegab_2\,, \delta \wb_2)$, and

\begin{equation}\label{C:slider}
\Cb = 
\left(
\begin{array}{cccc}
 \scriptstyle -(\pb_2 \times  \mb_1)^T  & \scriptstyle -  \mb_1^T &  \scriptstyle (\pb_2 \times  \mb_1)^T  &  \scriptstyle \mb_1^T \\ 
 \scriptstyle -(\pb_2 \times  \nb_1)^T  & \scriptstyle -  \nb_1^T &  \scriptstyle (\pb_2 \times  \nb_1)^T  &  \scriptstyle \nb_1^T 
\end{array}
\right)\,. \nonumber
\end{equation}
%



\subsubsection{Sliding hinge as a double point slider}
With the notation introduced previously, the constraint 
 imposes that the points $\pb_2$ and $\qb_1$ be the position vector of a point of $r_1$ and $r_2$, respectively;  
then 
\begin{eqnarray}
\pb_2 - \pb_1 \quad  || \quad  \ r_1 \,, \\ 
\qb_1 - \qb_2 \quad  || \quad  \ r_2 \,. \\ 
\end{eqnarray}
It follows that the constraint equations are
\begin{eqnarray}
c_1(\Rb_1\,, \ub_1\,, \Rb_2\,, \ub_2)  &=&  (\pb_2 - \pb_1)\cdot\mb_1  \,,   \nonumber\\
c_2(\Rb_1\,, \ub_1\,, \Rb_2\,, \ub_2)  &=&  (\pb_2 - \pb_1)\cdot\nb_1  \,,    \nonumber \\
c_3(\Rb_1\,, \ub_1\,, \Rb_2\,, \ub_2)  &=&  (\qb_1 - \qb_2)\cdot\mb_2  \,,   \nonumber\\
c_4(\Rb_1\,, \ub_1\,, \Rb_2\,, \ub_2)  &=&  (\qb_1 - \qb_2)\cdot\nb_2  \,,    \nonumber
\end{eqnarray}
with $\mb_i = \Rb_i \mb$ and  $\nb_i = \Rb_i \nb$, $i=1,2$.
%
%
%
%
The time derivatives of these  functions have the expressions 
\begin{eqnarray}
\dot c_1 &=&    \pb_2 \times \mb_1 \cdot  (\omegab_2 - \omegab_1) +  \mb_1 \cdot (\wb_2 - \wb_1)\,,   \nonumber \\
\dot c_2 &=&    \pb_2 \times \nb_1 \cdot  (\omegab_2 - \omegab_1) +  \nb_1 \cdot (\wb_2 - \wb_1)\,,   \nonumber  \\
\dot c_3 &=&    \qb_1 \times \mb_2 \cdot  (\omegab_1 - \omegab_2) +  \mb_2 \cdot (\wb_1 - \wb_2)\,,   \nonumber \\
\dot c_4 &=&    \qb_1 \times \nb_2 \cdot  (\omegab_1 - \omegab_2) +  \nb_2 \cdot (\wb_1 - \wb_2)\,.   \nonumber 
\end{eqnarray}
Then, the first order approximation of the  the constraint equations  in 
 the neighborhood of the configuration $(\Rb_1 \,, \ub_1\,,\Rb_2 \,, \ub_2)$,  has the expression
\begin{equation}
\cb_0 + \Cb \deltab = 0 \,,
\end{equation}
with $\cb = (c_1\,, c_2\,, c_3\,, c_4)$,   $\cb_0 = \cb(\Rb_1 \,, \ub_1\,,\Rb_2 \,, \ub_2)$, 
$\deltab = (\delta \omegab_1\,, \delta \wb_1\,, \delta \omegab_2\,, \delta \wb_2)$, and

\begin{equation}\label{p:slid:C}
\Cb = 
\left(
\begin{array}{cccc}
 \scriptstyle -(\pb_2 \times  \mb_1)^T  & \scriptstyle -  \mb_1^T &  \scriptstyle (\pb_2 \times  \mb_1)^T  &  \scriptstyle \mb_1^T \\ 
 \scriptstyle -(\pb_2 \times  \nb_1)^T  & \scriptstyle -  \nb_1^T &  \scriptstyle (\pb_2 \times  \nb_1)^T  &  \scriptstyle \nb_1^T    \\
 \scriptstyle -(\qb_1 \times  \mb_2)^T  & \scriptstyle -  \mb_2^T &  \scriptstyle (\qb_1 \times  \mb_2)^T  &  \scriptstyle \mb_2^T \\ 
 \scriptstyle -(\qb_1 \times  \nb_2)^T  & \scriptstyle -  \nb_2^T &  \scriptstyle (\qb_1 \times  \nb_2)^T  &  \scriptstyle \nb_2^T 

\end{array}
\right)\,. 
\end{equation}
\subsubsection{Spherical hinge}

For  a spherical  hinge we start with the conditions
$$
\Rb_1 \pb  + \ub_1=\Rb_2  \pb  +\ub_2\,,
$$
or, with the notation used in the previous section,
$$
 \pb_2 - \pb_1=\0\,.
$$
Then, the constraint function is 
$$
\cb = (\Rb_1- \Rb_2) \pb + \ub_1 - \ub_2  \,, \nonumber
$$
with  derivatives with respect to time 
$$
\dot\cb    =   \Wb_1  \pb_1   - \Wb_2 \pb_2 +   \wb_1 - \wb_2 \,.   \nonumber
$$
or, in terms of angular velocities, 
$$
\dot\cb    =  - \pb_1 \times   \omegab_1  + \pb_2 \times   \omegab_2  + \wb_1 - \wb_2\,. \nonumber
$$
Thus, the corresponding form of the $\Cb$ operator is
\begin{equation}\label{C:spherical}
\Cb = 
\left(
\begin{array}{cccc}
 - *\pb_1  & \ \ \Ib & \ \  *\pb_2  & \ \  -\Ib  
\end{array}
\right)\,. 
\end{equation}


\subsubsection{Door hinge as a point slider and a spherical hinge}

The operators corresponding to a point slider and to a spherical hinge can be composed together to give that of a door hinge: 
\begin{equation}\label{C:point:door}
\Cb = 
\left(
\begin{array}{cccc}
 \scriptstyle -(\qb_1 \times  \mb_2)^T  & \scriptstyle -  \mb_2^T &  \scriptstyle (\qb_1 \times  \mb_2)^T  &  \scriptstyle \mb_2^T \\ 
 \scriptstyle -(\qb_1 \times  \nb_2)^T  & \scriptstyle -  \nb_2^T &  \scriptstyle (\qb_1 \times  \nb_2)^T  &  \scriptstyle \nb_2^T  \\
 \scriptstyle  - *\pb_1  & \scriptstyle\Ib &  \scriptstyle *\pb_2  &  \scriptstyle -\Ib  \\
\end{array}
\right)\,. 
\end{equation}






\subsection{Integration of the kinematic-compatibility equations}


Let $E$ be  the Euclidean group. The elements of $E$ are the pair $(\Rb,\ub)$ with $\Rb\in \Orth^+$
 the rotation and $\ub\in V$  the translation.  The group operation on $E$ is defined by
 \begin{equation}
 (\Rb, \ub)\circ (\Rb',\ub')  =  (\Rb \Rb', \Rb \ub' + \ub)  \,,
 \end{equation}
 the inverse of $(\Rb,\ub)$ is  $(\Rb^{-1}, -\Rb\ub)$,  and the unit element of the group
 is the pair $(I, \0)$. This make $E$ the semidirect product of $\Orth^+$ and $V$.
 
 The Lie algebra $\eef$ of $E$  is given by the semi-direct product of the Lie algebras 
 $\Skw$ and  $V$.
 The exponential map  $\exp: \eef \to E$ transforms the pair $(\Wb,\wb)\in \eef$ in the pair  $(\Rb, \ub)\in E$, with
 \begin{equation}
  \Rb = \Ib + \sin(\theta)  \Wb + \frac{ (1-\cos(\theta)) }{   \theta^2} \Wb^2 \,,
 \end{equation}
  \begin{equation}
           \ub  =  \Ab \wb \,,
\end{equation}
 here  $\theta$ denotes the norm of the axial $\omegab$ of $\Wb$ and  $\Ab$ the operator
 \begin{equation}
           \Ab =       \Ib+  \frac{(1-\cos(\theta)) }{\theta^2}  \Wb + \frac{1}{\theta^2} (1- \frac{\sin(\theta)}{\theta})  \Wb^2 \,.
\end{equation}

Let $G$ be the direct product group of $N$ copies of $E$, that is $G =  \overbrace{E\times\cdots\times E}^{  N-\mathrm{times}}$.
In components, the elements g of $G$ are written 
  $g = (g_1, g_2, \dots, g_N)$,  with     $ g_i = (\Rb_i , \ub_i) \in E$.  The $i$-th component $g_i$  of $g$  defines the rigid transport  of the $i$-th body. 
Then the Lie algebra $\gef$ of $G$ is the direct product of $n$ copies of the Lie algebra $\eef$ of $E$, and
   the exponential map  $\exp: \gef \to G$  is the map that  transforms $ (\ub_1, \ub_1, \dots, \ub_N) \in \gef$ in 
    $g = (g_1, g_2, \dots, g_N)\in G$,  with $g_i$ the image of $\ub_i$ under the exponential map of the Euclidean group $E$. 
 

\subsubsection{Newton-Raphson  method on $\Real^n$}
To introduce the notation adopted, we briefly recall here the well-known Newton-Raphson method on $\Real^n$. Given  a map $f: \Real^n \rightarrow \Real^m$, we consider the problem of solving the nonlinear equation
\begin{equation}
  f(x) = 0  \,.
\end{equation}
The iteration of the Newton-Raphson method for solving this equation  is defined by
\begin{equation}
 f(x_n) + \nabla f (x_n) \delta x = 0 \,,
 \end{equation}   
 \begin{equation}
    x_{n+1} =  x_{n} + \delta x \,.
 \end{equation}   
 
Let  $p_{x_n}$ the function defined by 
$
p_{x_n} (\vb) = x_n + \vb\,,
$
and let   $ \hat f$ the composed function $ \hat f  = f \circ  p_{x_n} $.
 Then   
   \begin{equation} 
\hat f(\vb) = f(x_n+\vb ) \,,\quad\quad   \nabla \hat f(\vb) = \nabla f(x_n+\vb) \nabla p_{x_n}(\vb) =  \nabla f(x_n+\vb) \,,
 \end{equation}   
as  $ \nabla  p_{x_n} = I  $; in particular,  for $\vb = 0$, we have
   \begin{equation} 
\hat f(0) = f(x_n ) \,,\quad\quad   \nabla \hat f(0) = \nabla f(x_n) \nabla p_{x_n}(0) =  \nabla f(x_n) \,.
 \end{equation}   
 It follows that the Newton-Raphson iteration can be rewritten in the form
 \begin{equation}
\hat f(0) + \nabla f (x_n) \nabla p_{x_n}(0)  \delta x = 0 \,,
 \end{equation}   
 \begin{equation}
    x_{n+1} =  p_{x_n}(\delta x) \,.
 \end{equation}

 \subsubsection{Newton-Raphson Method on Lie groups}
 Let $G$ be a Lie group. For every $h\in G$, the right translation   is the map $R_{h} :G \to G$ defined by 
\begin{equation}
                     R_{g_n}(h)  = h \, g_n \,.
\end{equation}

Given  a map $f: G \rightarrow \Real^m$, we consider the problem of solving the nonlinear equation
\begin{equation}
  f(x) = 0  \,,
\end{equation}
in which $x$ is the collection of rotations and translations of all bodies, $(\Rb_i,\ub_i)$, $i~=~1,\ldots,N$, and $f$ is collection of all constraint functions $\cb$ that we detailed in the previous subsections for each type of constraint.

Let   $ \hat f$ be the composed function $ \hat f  = f \circ  R_{g_n}\circ \exp(\vb) $,
 with $\exp : \gef \to G$  the exponential map.
 Then   $\hat f(\vb) = f( \exp(\vb)\, g_n)$ and
   \begin{equation} 
   \nabla \hat f(\vb) = \nabla f(\exp(\vb) g_n) \circ \nabla R_{g_n}(\exp(\vb)\circ \nabla\exp(\vb)  \,;
 \end{equation}   
because the exponential of $0\in \gef$ is the identity $\eb \in G$, and the differential of the exponential map at zero is the identity  of $\gef$,
\begin{equation}
  \exp(0) = \eb  \,,\quad\quad  \nabla\exp(0)  = I\,,
\end{equation}
we have
   \begin{equation} 
\hat f(0) = f(g_n ) \,,\quad\quad   \nabla \hat f(0) = \nabla f(g_n) \circ \nabla R_{g_n}(\eb) \,.
 \end{equation}   

 It follows the Newton-Raphson iteration  on Lie group
 \begin{equation}
\hat f(0) + \nabla f (g_n) \circ  \nabla R_{g_n}   \delta \vb = 0 \,,
 \end{equation}   
 \begin{equation}
    g_{n+1} =  R_{g_n} \circ \exp (\delta \vb)  = \exp (\delta \vb) g_n\,,
 \end{equation}  
in which $\delta\vb$ is the collection of all the increments $(\delta\omegab_i,\delta\wb_i)$, $i~=~1,\ldots,N$. 

\subsubsection{Folding example}\label{sec:ykin}

\begin{figure}[h!]
\centering
\includegraphics[width=0.9\textwidth]{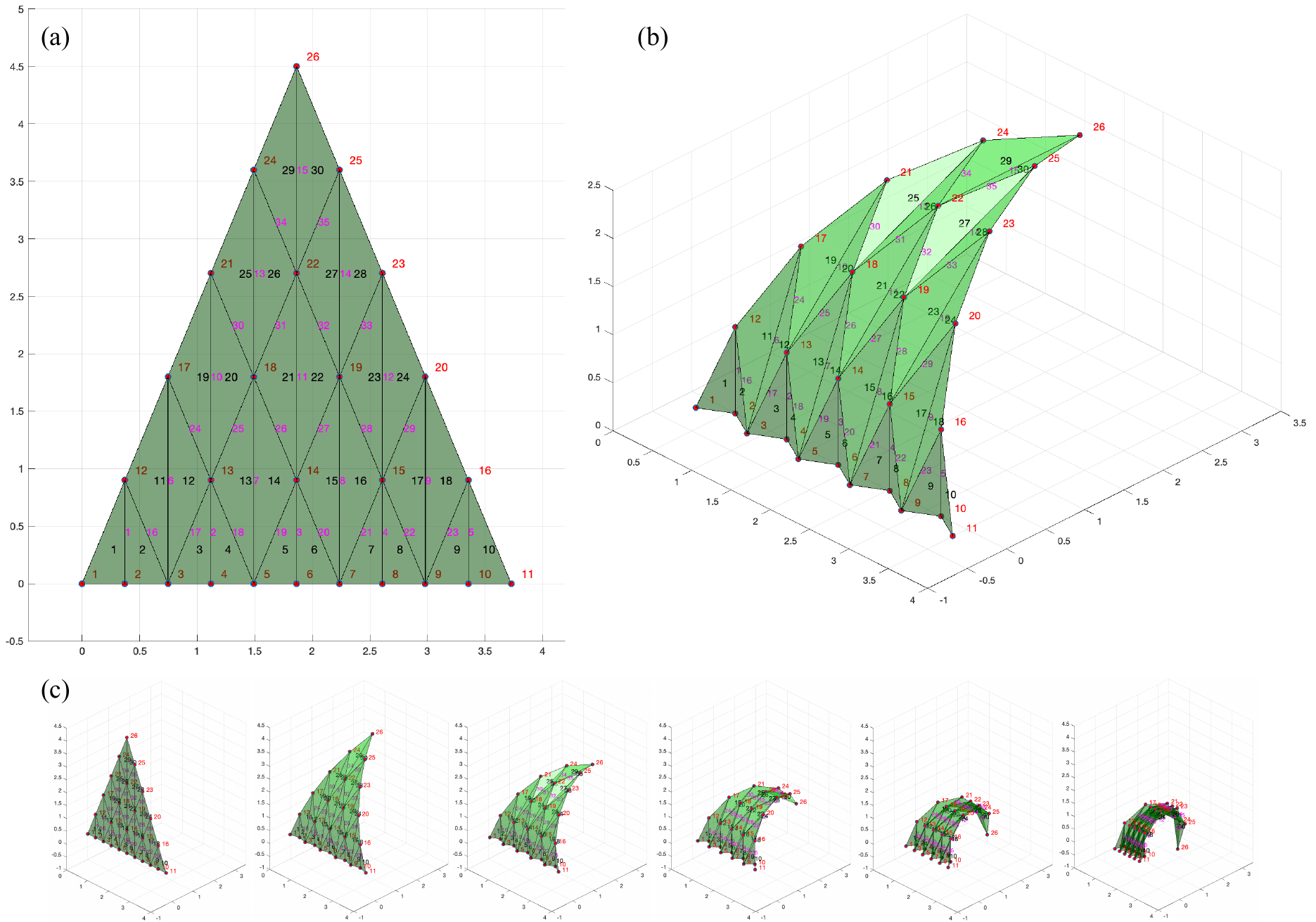}
\caption{Folding of a triangular portion of Yoshimura origami. (a) Initial configuration in the $x-z$ plane. (b) Partially folded configuration. (c) Snapshot of the folding process.}\label{yoshi}
\end{figure}

As the present work is mainly focused on the study of isostatic thick origami structures, we present just one folding example. We apply the above-described procedure to fold the triangular portion of Yoshimura pattern shown in Figure~\ref{yoshi}\,(a).

This triangulated origami structure is composed by isosceles triangles with a long side of $1.8$ units and two equal angles of $22.5$ degrees. The assembly has $P=30$ panels with $H=35$ hinge lines. The number of edges on the boundary is $E_b=20$. Hence, the number of internal independent mechanisms is $M=E_b-3=17$. We consider all hinges to be door hinges. The folding process is prescribed by imposing the relative angular velocities at the $15$ vertical hinges to be equal to $0.005$\,rad/s, while maintaining vertex $6$ fixed in space, the vertices $2, 4, 8, 10, 12, 13, 14, 15, 16$ moving only along the $x$ axis, and the motion of vertex $14$ blocked along the $x$ and $y$ direction.

Figure~\ref{yoshi}\,(b) shows the configuration reached after $200$\,s of simulation time. Figure~\ref{yoshi}\,(c) shows snapshots of the folding process taken every $100$\,s of simulation time.


\section{Static analyses} 

In this section, we first list the internal constraint reactions for each type of constraint, and then we report our analyses regarding the Miura-ori, the Yoshimura, and the Kresling crease patterns.

\subsection{Internal constraint reactions}

We recall that in Section \ref{sec:kin}, the first-order approximation of the internal constraint equation were obtained in the form
\begin{equation}
\cb_0 + \Cb \deltab = 0 \,, \nonumber
\end{equation}
with $\cb_0$ the constraint functions, $\deltab$ the increment of angular and translational velocities, and $\Cb$ an operator whose form depend on the considered constraint.

By assuming the constraints to be ideal,  the vector $\rb= ( \mb_1\,, \fb_1\,,  \mb_2\,, \fb_2)$ of the  reactive action at hinges, with $\mb_i,\,\fb_i$, $i=1,2$, the internal couple and the internal force exchanged at a hinge and reduced to the origin of the reference frame, is given by
\begin{equation}
\rb = \Bb \lambdab \,,
\end{equation}
with $\lambdab$  
the vector of internal actions expressed as Lagrange multipliers and
$\Bb = \Cb^T$. We detail below the form of the $\Bb$ operator for each constraint type.


For a sliding hinge, from \eqref{slid:C}, we have $\lambdab= (\lambda_1\,, \lambda_2\,, \lambda_3\,, \lambda_4)$ and 
\begin{equation}
\Bb =  \frac{1}{2}
\left(
\begin{array}{cccc}
 \scriptstyle - (\tb_1 \times \mb_2 + \tb_2 \times \mb_1)  &  \scriptstyle -(\tb_1 \times \nb_2 + \tb_2 \times \nb_1) 
      & \scriptstyle- ( \pb_2 \times \mb_1 + \pb_1 \times \mb_2)      & \scriptstyle- ( \pb_2  \times \nb_1 + \pb_1 \times \nb_2)   \\
\scriptstyle 0                                                             &   \scriptstyle 0                                                            
    & \scriptstyle -(\mb_1+ \mb_2)                                                & \scriptstyle- ( \nb_1+ \nb_2)                                             \\
\scriptstyle  \tb_1 \times  \mb_2 + \tb_2 \times  \mb_1 &  \scriptstyle \tb_1 \times \nb_2  + \tb_2  \times  \nb_1 
    & \scriptstyle  \pb_2  \times \mb_1 + \pb_1 \times \mb_2 & \scriptstyle   \pb_2  \times \nb_1 + \pb_1 \times \nb_2   \\
\scriptstyle 0                                                            & \scriptstyle 0                                                             
    & \scriptstyle\mb_1+ \mb_2                                                  &   \scriptstyle  \nb_1+ \nb_2                                          \\
\end{array}
\right)\,; \nonumber
\end{equation}
for a sliding hinge as a double point slider, from \eqref{p:slid:C}, we have instead 
\begin{equation}
\Bb = 
\left(
\begin{array}{cccc}
    \scriptstyle -(\pb_2 \times  \mb_1)  & \scriptstyle -(\pb_2 \times  \nb_1)   & \scriptstyle -(\qb_1 \times  \mb_2)  & \scriptstyle -(\qb_1 \times  \nb_2) \nonumber \\ 
    \scriptstyle -  \mb_1                     & \scriptstyle -  \nb_1                       & \scriptstyle -  \mb_2                     & \scriptstyle -  \nb_2                     \nonumber \\ 
    \scriptstyle (\pb_2 \times  \mb_1)   &  \scriptstyle (\pb_2 \times  \nb_1)   & \scriptstyle (\qb_1 \times  \mb_2)   &  \scriptstyle (\qb_1 \times  \nb_2)   \nonumber \\ 
    \scriptstyle \mb_1                        &  \scriptstyle \nb_1                         &  \scriptstyle \mb_2                        &  \scriptstyle \nb_2
\end{array}
\right)\,.
\end{equation}
%
For a door hinge, from \eqref{door:C}, we have $\lambdab = (\lambda_1\,, \lambda_2\,, \lambdab_3)$ and 
\begin{equation}
\Bb =  \frac{1}{2}
\left(
\begin{array}{cccc} 
 \scriptstyle - (\tb_1 \times \mb_2 + \tb_2 \times \mb_1)  &  \scriptstyle - (\tb_1 \times \nb_2 + \tb_2 \times \nb_1)  &\scriptstyle -\,*\pb_1  \\
\scriptstyle 0                                                             &  \scriptstyle 0                                                            &\scriptstyle  -\Ib
\\
\scriptstyle  \tb_1 \times  \mb_2 + \tb_2 \times  \mb_1 &  \scriptstyle \tb_1 \times \nb_2  + \tb_2  \times  \nb_1       &\scriptstyle *\pb_2  \\
\scriptstyle 0                                                            & \scriptstyle 0                                                              &\scriptstyle  \Ib
\\
\end{array}
\right)\,; \nonumber
\end{equation}
%
%
%
%
for a door hinge as a point slider and a spherical hinge, from \eqref{C:point:door}, we have instead
\begin{equation}\label{B:point:door}
\Bb = 
\left(
\begin{array}{cccc}
  \scriptstyle -(\qb_1 \times  \mb_2)  & \scriptstyle -(\qb_1 \times  \nb_2) \nonumber &  \scriptstyle  *\pb_1    \\ 
 \scriptstyle -  \mb_2                     & \scriptstyle -  \nb_2 &  \scriptstyle\Ib    \\
 \scriptstyle (\qb_1 \times  \mb_2)   &  \scriptstyle (\qb_1 \times  \nb_2) &  \scriptstyle -\,*\pb_2     \\
 \scriptstyle \mb_2                        &  \scriptstyle \nb_2 & \scriptstyle -\Ib \\
\end{array}
\right)\,. 
\end{equation}

The equilibrium equations of the whole origami structure are obtained by imposing, for each body, the resultant of all the internal constraint reaction and the external loads applied to it, and the resultant moment of the same with respect to a fixed pole, to be null.


\subsection{Examples}

\begin{figure}[t!]
\centering
\includegraphics[width=0.9\textwidth]{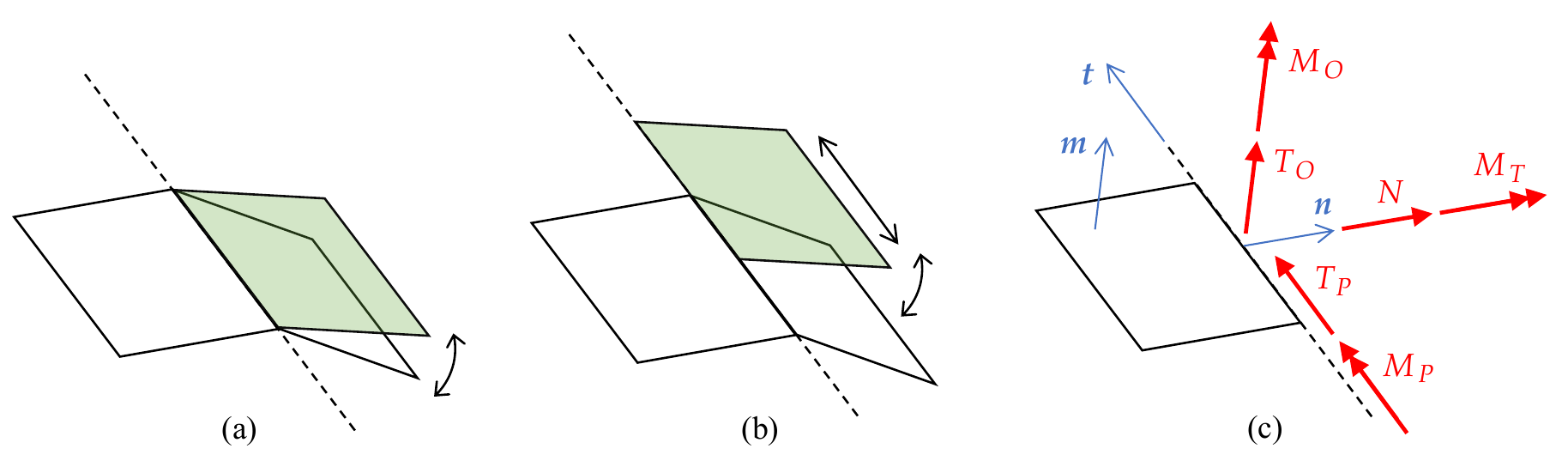}
\caption{Two panels connected by a door hinge (a) and by a sliding hinge (b). Representations of the internal actions applied to a panel edge. The parallel moment $M_P$ is always null; the parallel shear force $T_P$ is null for sliding hinges.}\label{int_act}
\end{figure}

In order to report the results of the static analyses, we reduce the constraints reactions to the mid points of the corresponding edges shared by adjacent panels, and project them along the directions of a right-handed local frame $\{\nb,\tb,\mb\}$ attached to one of the two panels, with $\nb$ the outward normal to the edge in the panel plane, $\tb$ the tangent to the edge, and $\mb$ the normal to the panel. In this way, the internal force and couple applied to a panel by the adjacent one are described by the triplets $(N, T_P,T_O)$ and $(M_T,M_P,M_O)$, respectively, with $N$ the normal force, $T_P$ the parallel shear force, $T_O$ the orthogonal shear force, $M_T$ the twisting moment, $M_P$ the parallel moment, and $M_O$ the orthogonal moment, as illustrated in Figure~\ref{int_act}. The parallel moment is always null, while the parallel shear force is null for sliding hinges.

\begin{figure}[t!]
\centering
\includegraphics[width=0.9\textwidth]{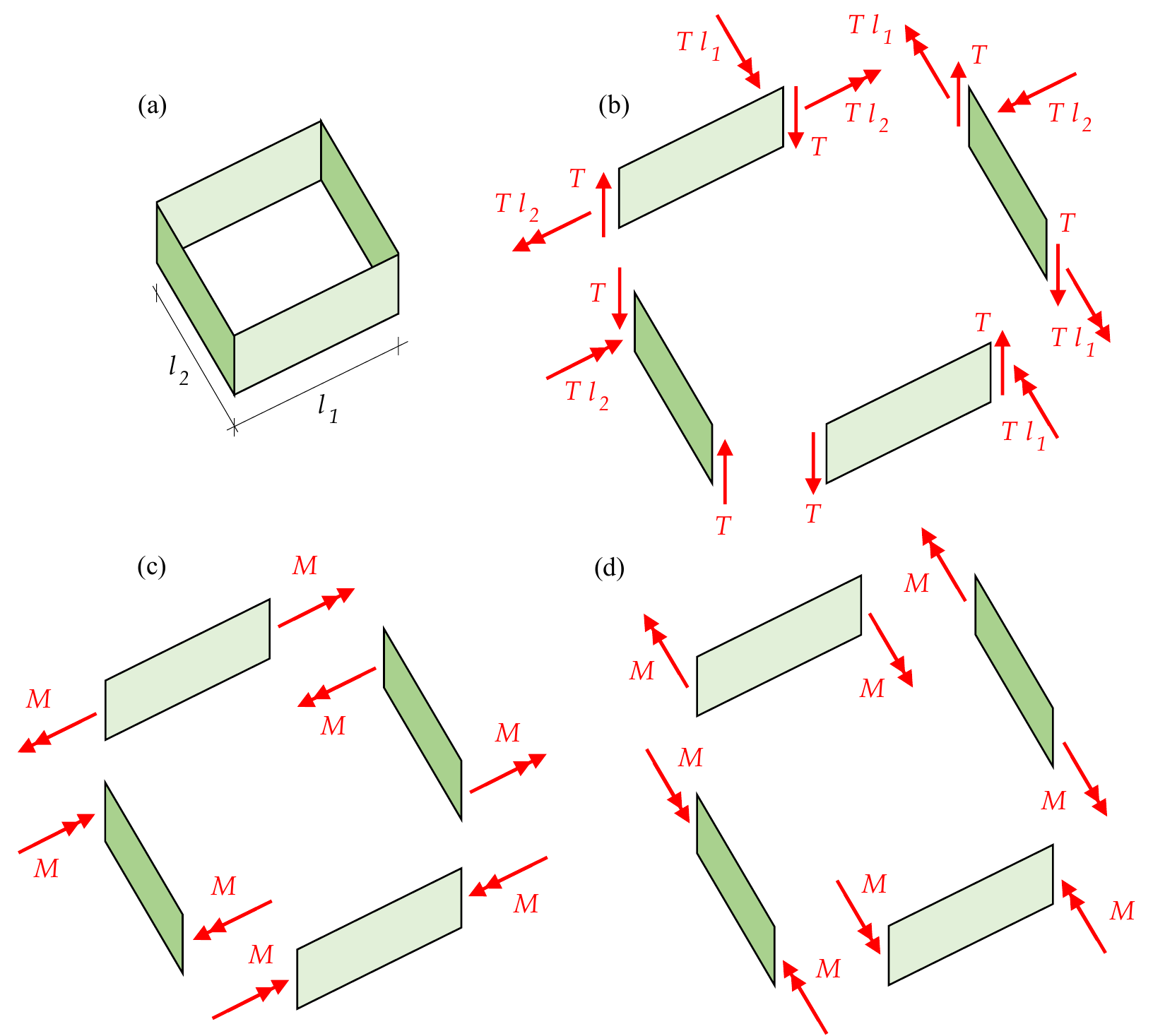}
\caption{A four-panel origami ring (a) and a representation of its three self-stress states (b,c,d).}\label{belt}
\end{figure}

We introduce our gallery of examples by describing the simple case of a panel ring formed by four rectangular panels connected by four door hinges, which is shown in Figure~\ref{belt}~(a).
The kinematics of this assembly is analogous to that of a planar four-bar linkage and therefore there is just one internal mechanisms, $M=1$. Since there are no internal vertices, the number of self-stress states is $S=3$, according to both the bar-framework model and the panel-hinge model. 
In addition, it is easy to check by straightforward equilibrium calculations that the internal actions $N$ and $T_O$ must be null for all panel hinges. Then, by taking as parameters the parallel shear force, the twisting moment, and the orthogonal moment at one hinge, one can compute explicitly the internal actions for the three self-stress states, which are represented in Figure~\ref{belt}~(b,c,d).
It is easy to check that panel rings with more than four panels, $P>4$, always have $S=3$ and $M=P-3$.

It is worth noticing that by replacing door hinges by sliding hinges it is only possible to eliminate the self-stress state with nonzero parallel shear $T$ (Figure~\ref{belt}~(b)). The introduction of more than one sliding hinge would generate additional internal mechanisms without affecting the other two independent self-stress states.

\subsubsection{Miura-ori patches}

\begin{figure}[t!]
\centering
\includegraphics[width=0.35\textwidth]{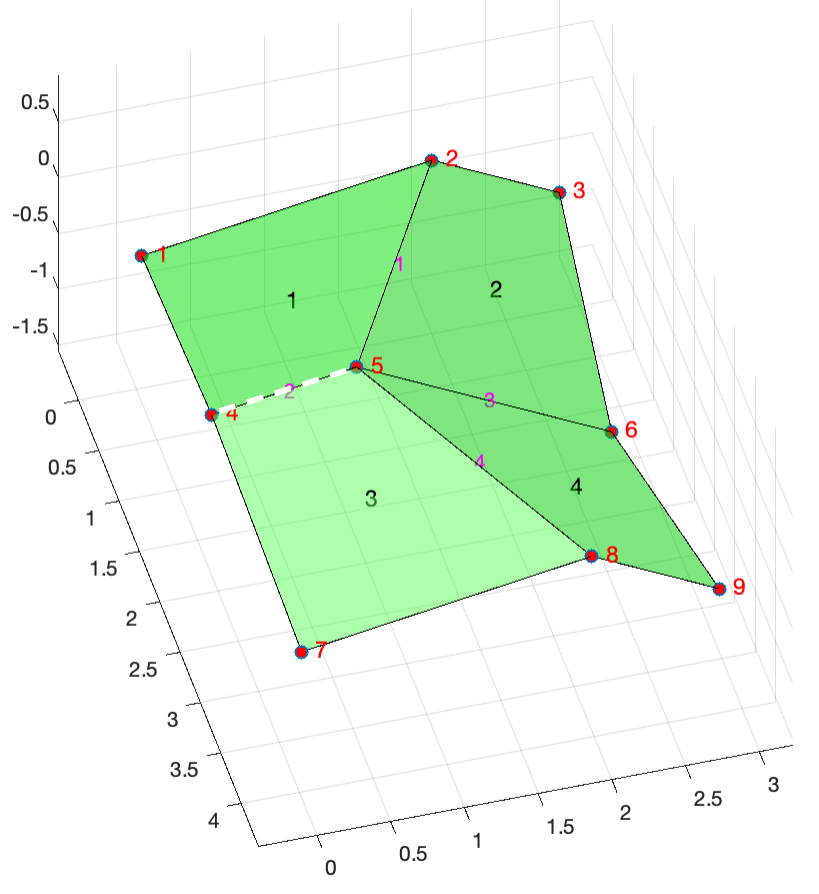}\ \ \
\includegraphics[width=0.61\textwidth]{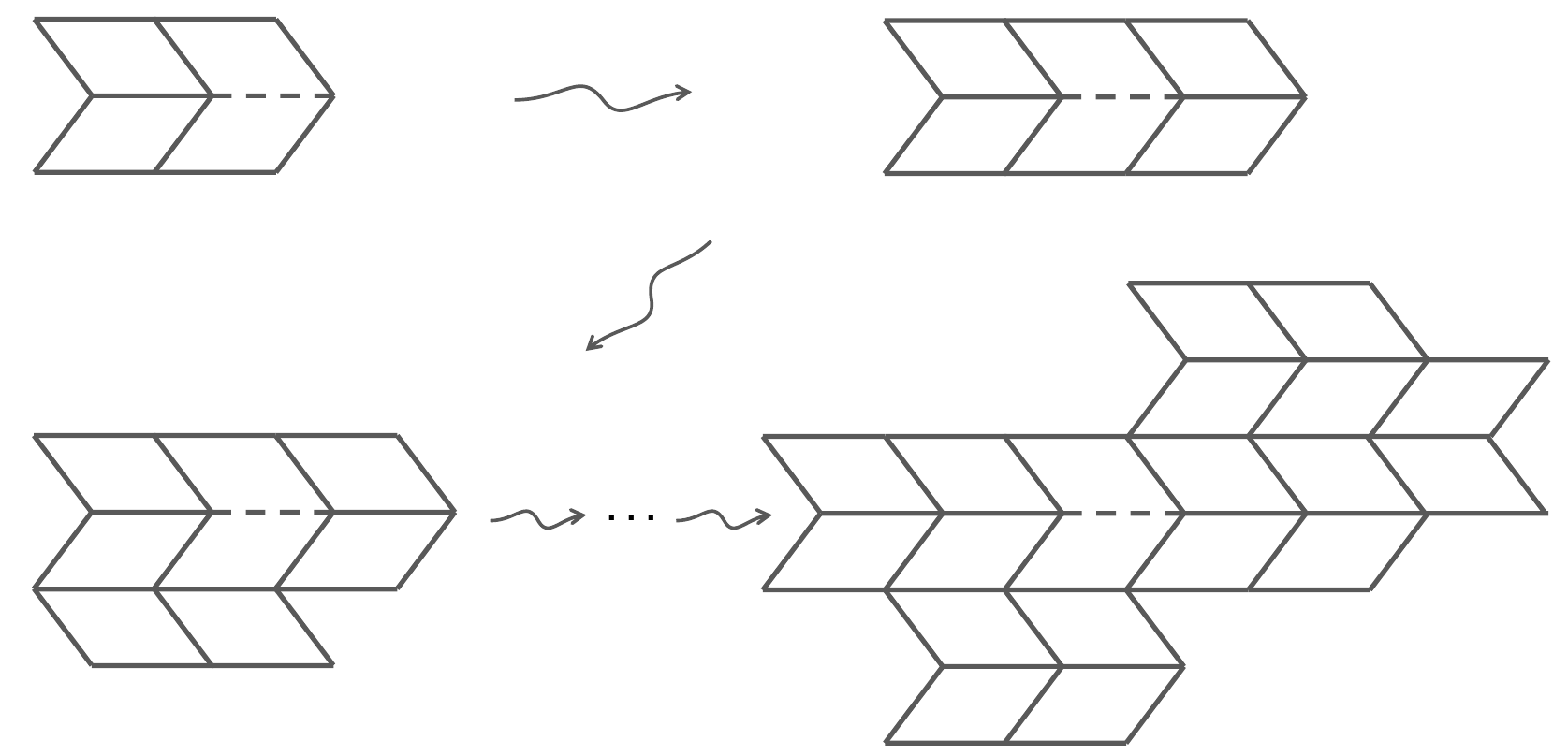}
\caption{Left: a single-vertex Miura-ori origami with null degree of static indeterminacy. There are three sliding hinges and one door hinge (dashed line). Right: sequential assembly of isostatic Miura-ori origami. There is just a single door hinge (dashed line), all the other hinges are sliding hinges. At each step, two panels and three sliding hinges are added.}\label{miura}
\end{figure}

Similar to the case of the four-panel ring, the single-vertex Miura-ori origami shown in Figure~\ref{miura}\,(left) has $M=1$ and $S_{PH}=3$, when all the four hinges are door hinges. However, by performing numerical calculations with the proposed formulation, we checked that at non-singular configurations the replacement of any three door hinges with three sliding hinges makes the degree of static indeterminacy equal to zero. 

\begin{figure}[h]
\centering
\includegraphics[width=0.8\textwidth]{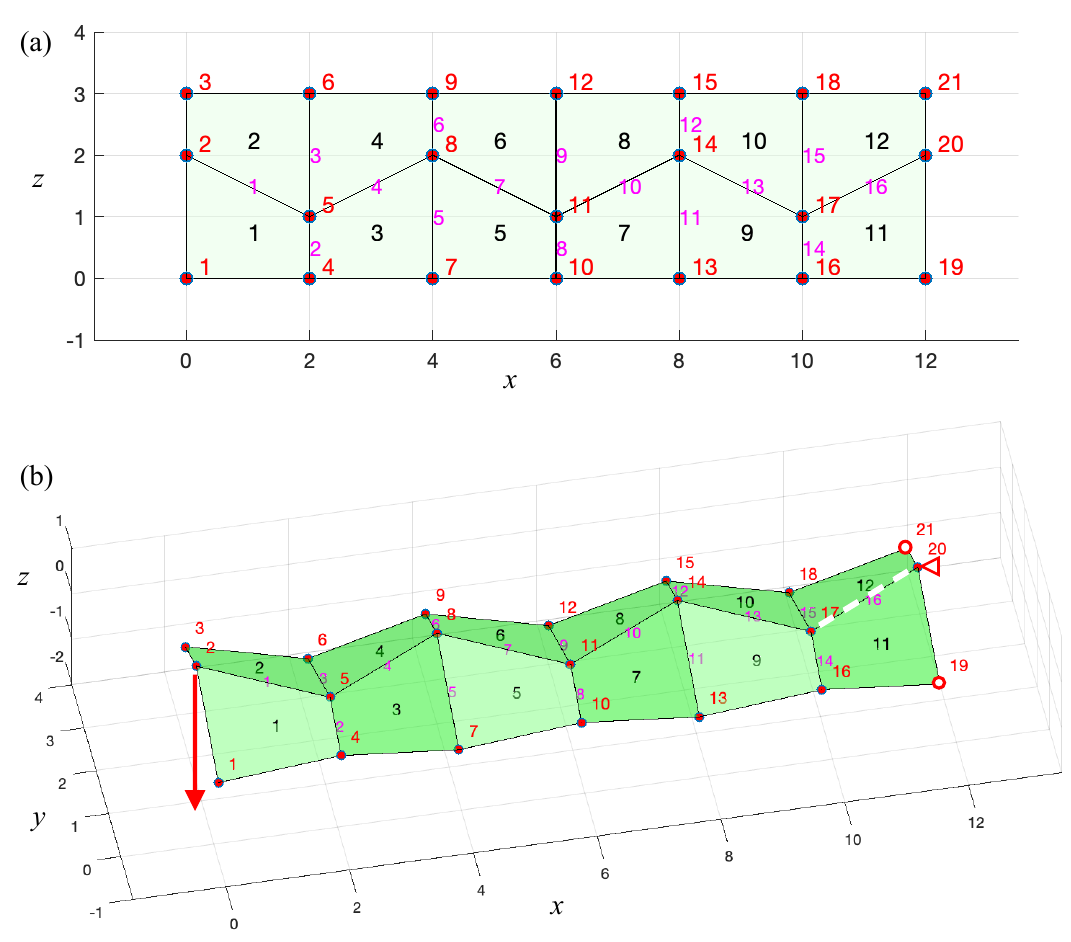}
\caption{An isostatic Miura-ori cantilever loaded at the tip. Flat configuration before folding (a). Cantilevered structure (b).}\label{canti}
\end{figure}
%
%
%
Regular Miura-ori origami structures always possess just one internal mechanisms, no matter the number of panels they are composed of. Therefore, the degree of static indeterminacy always increases with the size of the Miura-ori patch. We found in our analyses that that there is a class of Miura-ori patches whose degree of static indeterminacy can be made equal to zero. The elements of this class can be constructed sequentially, by starting from a single-vertex Miura-ori analogous to the one just described, and by iteratively adding two panels and three sliding hinges, as indicated in Figure~\ref{miura}\,(right).

%
%
%
We now give an example of calculation of internal actions for the case of the cantilevered Miura-ori beam shown in Figure~\ref{canti}. Each panel has the shape of a right trapezoid with height of $2$ units, and bases of $1$ and $2$ units. The configuration in Figure~\ref{canti}\,(b) is obtained by folding the initially flat configuration in Figure~\ref{canti}\,(a) so that the angle between the normals to panels $2$ and $3$ is equal to $27.4$~degrees. Table~\ref{tab:canti:ref} reports the corresponding vertex coordinates.
In the configuration of Figure~\ref{canti}\,(b), two corner vertices on the short side of the assembly are pinned to the ground
, and the middle vertex on the same side is constrained not to move along the longitudinal direction. 
At the other end of the assembly, a vertical point loads of unit magnitude is applied.
The assembly has one door hinge (dashed line in Figure~\ref{canti}), and the remaining hinges are sliding hinges. 
The resulting internal and external constraint reactions are shown in Table~\ref{tab:canti:int} and Table~\ref{tab:canti:ext}, respectively.

\subsubsection{Yoshimura wedges}

\begin{figure}[h]
\centering
\includegraphics[width=0.9\textwidth]{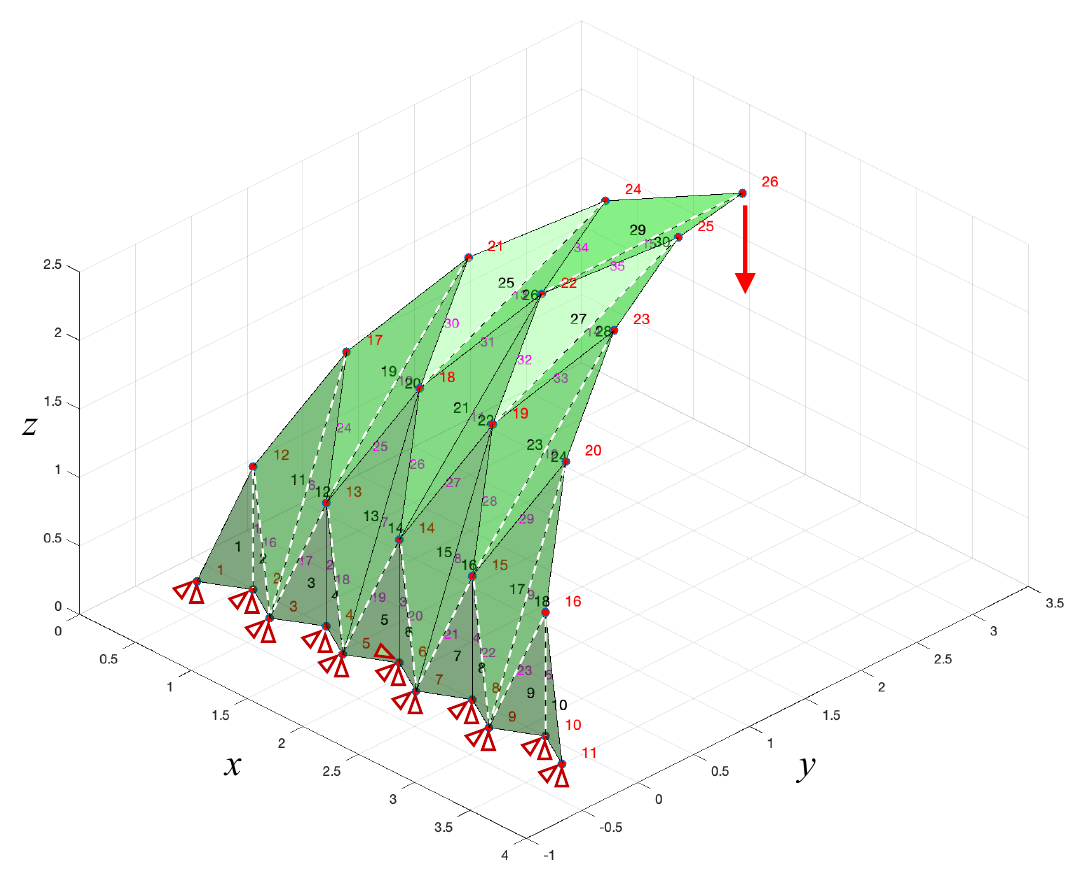}
\caption{A Yoshimura origami wedge shaped as a cantilevered arch loaded at the tip. Door hinges are denoted by dashed white  lines, the rest of them are sliding hinges.}\label{yoshi_load}
\end{figure}

We pass now to the Yoshimura origami structure described in Section~\ref{sec:ykin}. We consider the configuration shown in Figure~\ref{yoshi}\,(b) and reassign vertex constraints as follows. Al base vertices, those on the plane $z=0$, are fixed in the $y$ and $z$ direction, with vertex $6$ constrained also in the $x$ direction. 
We checked that with this set of vertex constraints, there are no internal mechanisms, $M=0$. This origami structure has 6 internal vertices. When all hinges are door hinges, the degree of static indeterminacy is $S_{PH} = 3  V_i = 18$. We verified that by replacing three door hinges at each internal vertex with three sliding hinges (cf. Figure~\ref{yoshi_load}) a isostatic structure is obtained.
We considered the case of a vertical unit load applied to the tip of the structure, as shown in Figure~\ref{yoshi_load}.
Table~\ref{tab:yoshi:ref} lists the nodal coordinates of the analyzed configuration, while Table~\ref{tab:yoshi:int} and Table~\ref{tab:yoshi:ext} reports the internal and external reactions, respectively.

We checked that it is possible to obtain similar isostatic Yoshimura wedges with any number of base elements in an anologous way by constraining all base vertices in the transversal ($y$) and vertical ($z$) directions, with one of them constrained also in the folding direction ($x$), and by replacing three door hinges with sliding hinges at each internal vertex.

\subsubsection{Kresling columns}

\begin{figure}[h]
\centering
\includegraphics[width=0.9\textwidth]{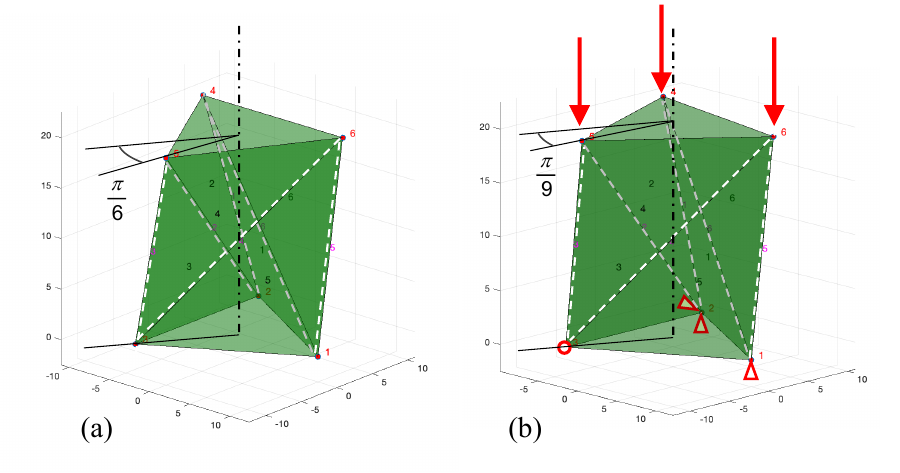}
\caption{Two different versions of a Kresling module with triangular base. All hinges are door hinges in both (dashed white lines). Configuration (a) is singular with one infinitesimal mechanism and one self-stress state; configuration (b) is isostatic. Configuration (b) is simply supported on the ground, and subjected to vertical unit forces.}\label{kresling3}
\end{figure}

We now pass to consider multimodular Kresling columns with triangular base.
Each module of a Kresling column is composed by $P=6$ triangular panels connected by $H=6$ door hinges in a three-fold cyclic-symmetric configuration, as shown in Figure~\ref{kresling3}. 
The bottom and top horizontal bases of a module appear rotated relatively to each other by a certain twist angle $\theta\in(\pi/3,\pi)$.  
We remark that given the panels of a Kresling module, it is possible to assemble them into two distinct configurations with opposite handedness, mirror image of each other. 

A one-module assembly has no internal vertices, and we found that  is always isostatic  ($M=0,\,S=0$), except for values of the twist angle equal to $\frac{\pi}{6}$, or  to $-\frac{5}{6}\pi$, at which the configuration is singular and admits one internal mechanism and one self-stress state ($M=1,\,S=1$). Figure~\ref{kresling3}\,(a) shows the configuration of a Kresling module with height $20$ units, bases inscribed in a circle with diameter of $20$ units, and twist angle $\theta=\frac{\pi}{6}$. The corresponding vertex coordinates are listed in Table~\ref{tab:kres:ind:ref}. We computed the values of the internal actions in the self-stress state reported in Table~\ref{tab:kres:ind:int}. 
Figure~\ref{kresling3}\,(b) shows a (isostatic) configuration with same height and base dimensions, and $\theta=\frac{\pi}{9}$, simply supported at the bottom base, and subjected to three vertical unit loads applied to the top vertices. Table~\ref{tab:kres:ref} lists the vertex coordinates, while Table~\ref{tab:kres:int} and Table~\ref{tab:kres:ext} reports the internal and external reactions, respectively.

\begin{figure}[h]
\centering
\includegraphics[width=0.8\textwidth]{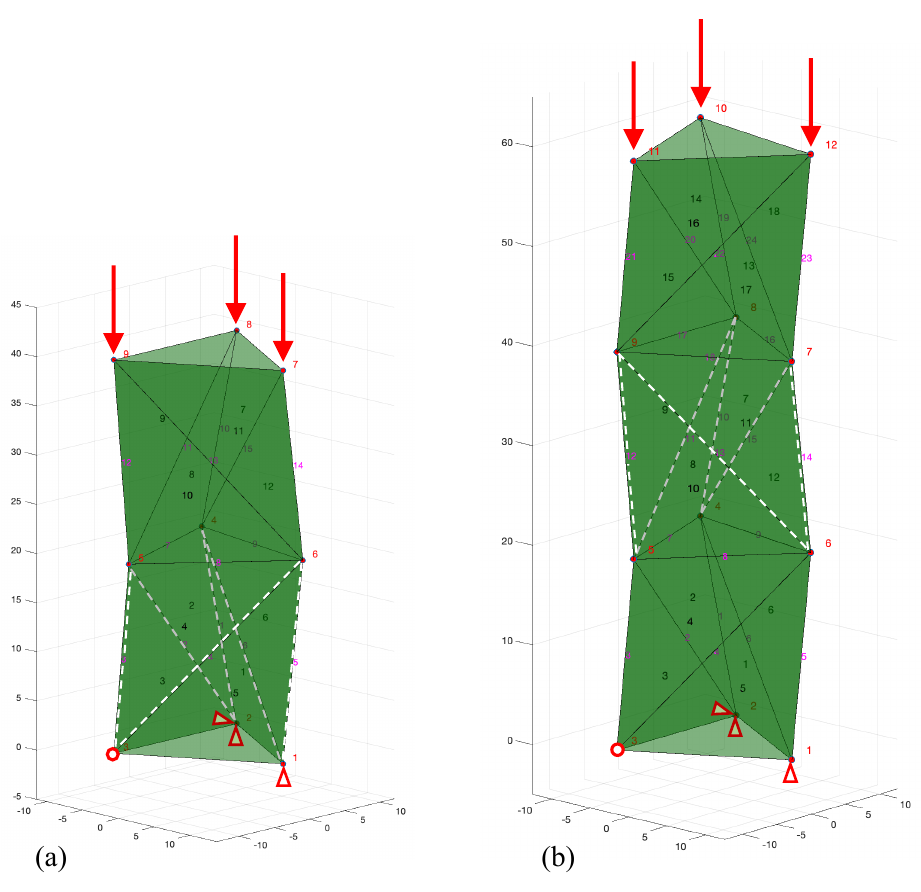}
\caption{Two isostatic Kresling towers with two (a) and three (b) modules. Six hinges on one of the modules are door hinges (dashed white lines).}\label{kresling3multi}
\end{figure}

A two-module column obtained by juxtaposition of two modules with same dimensions and opposite handedness is shown in Figure~\ref{kresling3multi}\,(a). In addition to the hinges belonging to each individual module, three more hinges realize the junction between the two module. This origami structure possesses three internal vertices and therefore has $S_{PH}=9$ degrees of static indeterminacy in non-singular configurations, while $M=0$. 
We checked that by realizing all hinges as sliding hinges, except for those belonging to one of the two modules, realized as door hinges, the assembly is isostatic. Isostatic multimodular Kresling columns can be obtained in a similar way. Figure~\ref{kresling3multi}\,(b) shows the case of a three-module tower. Tables~\ref{tab:k2} and \ref{tab:k3} shows the internal actions of the simply supported structures in Figure~\ref{kresling3multi}\,(a) and (b), respectively, resulting from a vertical loading by three unit forces on the top vertices.



\section{Conclusions}

We presented a design strategy aimed at obtaining  isostatic thick origami structures by making use of sliding hinges to replace conventional hinges. This strategy can effectively reduce the degree of static indeterminacy, and in many cases it can successfully make certain types of origami structures isostatic. 
We provided exact and linearized kinematic constraint equations and a numerical procedure to simulate the folding process by controlling the relative angular velocities at some creases of the origami.
Equilibrium equations relating the external loads to the internal actions exchanged by adjacent panels were obtained by duality.
We analyzed a series of noteworthy examples of isostatic thick origami structures. 
We showed that Miura-ori crease patterns in a certain class admit an isostatic realization. We found that Yoshimura wedges can always be made isostatic. Finally, we demonstrated isostatic multimodular Kresling origami.
The present method can be employed to discover, investigate, and design other types of isostatic thick origami structures and aid in the determination of self-stress states when this is not possible.

\backmatter



\bibliography{rigori}

\begin{table}
\caption{Nodal coordinates for the Miura-ori cantilever in Figure~\ref{canti}(b).}
\label{tab:canti:ref}
\centering
\small
\renewcommand{\arraystretch}{1.25}
\begin{tabular}{c l l l}
\hline\hline
\multicolumn{1}{c}{vertex} &
\multicolumn{1}{c}{$x$} &
\multicolumn{1}{c}{$y$} &
\multicolumn{1}{c}{$z$} \\
\hline
   1 &   1.1407e-01 &   1.9321e-01 & -8.576664e-01  \\
   2 &   1.1405e-01 &   2.0000e+00 & 1.298076e-04   \\
   3 &   1.1406e-01 &   2.9037e+00 & -4.283182e-01  \\
   4 &   2.0570e+00 &  -1.0121e-02 & -4.291978e-01  \\
   5 &   2.0570e+00 &   8.9329e-01 & 4.229811e-04  \\
   6 &   2.0571e+00 &   2.7007e+00 & -8.567885e-01  \\
   7 &   4.0000e+00 &   1.9317e-01 & -8.575398e-01  \\
   8 &   4.0000e+00 &   2.0000e+00 & 2.918110e-12  \\
   9 &   4.0000e+00 &   2.9037e+00 & -4.281852e-01  \\
  10 &   5.9430e+00 &  -1.0162e-02 & -4.290680e-01  \\
  11 &   5.9430e+00 &   8.9325e-01 & 2.931740e-04  \\
  12 &   5.9430e+00 &   2.7006e+00 & -8.566587e-01  \\
  13 &   7.8860e+00 &   1.9313e-01 & -8.574100e-01  \\
  14 &   7.8860e+00 &   2.0000e+00 & -1.298071e-04  \\
  15 &   7.8860e+00 &   2.9036e+00 & -4.280554e-01  \\
  16 &   9.8289e+00 &  -1.0203e-02 & -4.289382e-01  \\
  17 &   9.8289e+00 &   8.9321e-01 & 1.633668e-04  \\
  18 &   9.8290e+00 &   2.7006e+00 & -8.565289e-01  \\
  19 &   1.1772e+01 &   1.9309e-01 & -8.572739e-01  \\
  20 &   1.1772e+01 &   1.9999e+00 & -2.596132e-04  \\
  21 &   1.1772e+01 &   2.9036e+00 & -4.279319e-01  \\
\hline\hline
\end{tabular}
\normalsize
\end{table}

\begin{table}
\caption{Internal actions in the origami structure in Figure\,\ref{canti}\,(b).}
\label{tab:canti:int}
\centering
\small
\renewcommand{\arraystretch}{1.25}
\begin{tabular}{c c l l l l l l}
\hline\hline
\multicolumn{1}{c}{hinge} &
\multicolumn{1}{c}{panel} &
\multicolumn{1}{c}{$M_T$} &
\multicolumn{1}{c}{$M_P$} &
\multicolumn{1}{c}{$M_O$} &
\multicolumn{1}{c}{$N$} &
\multicolumn{1}{c}{$T_P$} &
\multicolumn{1}{c}{$T_O$} \\
\hline
     1    &     1     &    1.1261e+00 & -7.3184e-18  & 2.0638e+00  & 2.4004e-01  & 7.2964e-05 & -1.3053e-01  \\
     2    &      1    &     2.8615e-01 &  1.1102e-16 & -2.3319e+00 & -2.0083e-04 & -5.5511e-17 &  5.6919e-01  \\
     3    &      2    &    -5.7049e-01 &  3.3307e-16 &  2.3304e+00 & -8.2821e-06 & -5.5511e-17 &  5.7038e-01  \\
     4    &      3    &    -1.1309e+00 &  8.2345e-17 & -2.0748e+00 &  4.7603e-04 & -1.2365e-08 &  3.6941e-05  \\
     5    &      3    &     1.5807e+00 &  2.2204e-16 & -4.1433e+00 & -2.6260e-01 &  2.7756e-17 &  5.0514e-01  \\
     6    &      4    &     8.2005e-01 &  0.0000e+00 &  4.5364e+00 &  2.6263e-01 & -2.7756e-17 &  5.0596e-01  \\
     7    &      5    &     1.2698e+00 & -4.5103e-17 &  2.3295e+00 &  3.3092e-04 & -5.2939e-09 &  1.1381e-05  \\
     8    &      5    &    -7.8788e-01 &  4.4409e-16 & -6.7351e+00 & -5.2445e-04 &  2.7756e-17 &  5.6931e-01  \\
     9    &      6    &    -1.6445e+00 & -4.4409e-16 &  6.7323e+00 &  3.7505e-04 & -2.7756e-17 &  5.7034e-01  \\
    10    &      7    &    -1.1338e+00 &  1.0300e-18 & -2.0803e+00 &  4.0871e-04 &  2.2862e-08 & -7.9568e-05  \\
    11     &     7    &     2.6573e+00 & -1.3323e-15 & -8.5520e+00 & -2.6291e-01 & -2.7756e-17 &  5.0498e-01  \\
    12     &     8    &     1.8928e+00 &  4.4409e-16 &  8.9448e+00 &  2.6291e-01 & -2.7756e-17 &  5.0590e-01  \\
    13     &     9    &     1.2683e+00 & -3.1225e-17 &  2.3271e+00 &  2.6350e-04 &  2.6437e-08 & -7.1372e-05  \\
    14     &     9    &    -1.8630e+00 & -8.8818e-16 & -1.1141e+01 & -8.3424e-04 &  5.5511e-17 &  5.6939e-01  \\
    15     &    10    &    -2.7190e+00 &  0.0000e+00 &  1.1138e+01 &  7.4514e-04 & -2.7756e-17 &  5.7032e-01  \\
    16     &    11    &    -4.1820e+00 & -1.2295e-16 & -7.6601e+00 & -4.9841e+00 &  1.2588e+01 &  2.7246e+00  \\
\hline\hline
\end{tabular}
\normalsize
\end{table}

\begin{table}
\caption{Reaction forces at supports, and panels which they are applied to,  for the origami structure in Figure\,\ref{canti}\,(b).}
\label{tab:canti:ext}
\centering
\small
\renewcommand{\arraystretch}{1.25}
\begin{tabular}{c c l l l }
\hline\hline
\multicolumn{1}{c}{vertex} &
\multicolumn{1}{c}{panel} &
\multicolumn{1}{c}{$F_x$} &
\multicolumn{1}{c}{$F_y$} &
\multicolumn{1}{c}{$F_z$} \\
\hline
     11   &  19 &  -6.8092e+00 &  1.0571e+00 &  5.0081e-01 \\
     11   &  20 &  2.0423e+01 &  0.0000e+00 &  0.0000e+00 \\
     12   &  21 &  -1.3614e+01 & -1.0571e+00 &  4.9919e-01 \\
\hline\hline
\end{tabular}
\normalsize
\end{table}

\begin{table}
\caption{Nodal coordinates for the Yoshimura wedge in Figure~\ref{yoshi_load}.}
\label{tab:yoshi:ref}
\centering
\small
\renewcommand{\arraystretch}{1.25}
\begin{tabular}{c l l l}
\hline\hline
\multicolumn{1}{c}{vertex} &
\multicolumn{1}{c}{$x$} &
\multicolumn{1}{c}{$y$} &
\multicolumn{1}{c}{$z$} \\
\hline
   1 &  2.2806e-01 & -1.7867e-01 & 8.760393e-05  \\
   2 &  5.5537e-01 & -2.2498e-04 & 5.584535e-08  \\
   3 &  8.8252e-01 & -1.7896e-01 & 8.795992e-05  \\
   4 &  1.2096e+00 & -1.1246e-04 & -2.770419e-09  \\
   5 &  1.5368e+00 & -1.7874e-01 & 3.583811e-07  \\
   6 &  1.8640e+00 & -1.2845e-13 & 9.233695e-15  \\
   7 &  2.1911e+00 & -1.7874e-01 & 3.584869e-07  \\
   8 &  2.5182e+00 &  1.1253e-04 & 4.181614e-10  \\
   9 &  2.8454e+00 & -1.7853e-01 & -1.375840e-06  \\
  10 &  3.1725e+00 &  2.2495e-04 & 7.539000e-10  \\
  11 &  3.4996e+00 & -1.7872e-01 & -1.750372e-06  \\
  12 &  5.5537e-01 &  2.1689e-04 & 8.999997e-01  \\
  13 &  1.2097e+00 &  1.0814e-04 & 9.000000e-01  \\
  14 &  1.8640e+00 &  7.0238e-13 & 9.000000e-01  \\
  15 &  2.5182e+00 &  1.0816e-04 & 9.000000e-01  \\
  16 &  3.1725e+00 &  2.1686e-04 & 8.999997e-01  \\
  17 &  8.8263e-01 &  5.0949e-01 & 1.663227e+00  \\
  18 &  1.5369e+00 &  5.0929e-01 & 1.663310e+00  \\
  19 &  2.1910e+00 &  5.0929e-01 & 1.663314e+00  \\
  20 &  2.8453e+00 &  5.0915e-01 & 1.663456e+00  \\
  21 &  1.2099e+00 &  1.2717e+00 & 2.173996e+00  \\
  22 &  1.8640e+00 &  1.2718e+00 & 2.173735e+00  \\
  23 &  2.5182e+00 &  1.2714e+00 & 2.174276e+00  \\
  24 &  1.5368e+00 &  2.1716e+00 & 2.353834e+00  \\
  25 &  2.1915e+00 &  2.1713e+00 & 2.354493e+00  \\
  26 &  1.8647e+00  & 3.0718e+00 & 2.177855e+00  \\
\hline\hline
\end{tabular}
\normalsize
\end{table}

\begin{table}
\caption{Internal actions in the origami structure in Figure\,\ref{yoshi_load}.}
\label{tab:yoshi:int}
\centering
\small
\renewcommand{\arraystretch}{1.25}
\begin{tabular}{c c l l l l l l}
\hline\hline
\multicolumn{1}{c}{hinge} &
\multicolumn{1}{c}{panel} &
\multicolumn{1}{c}{$M_T$} &
\multicolumn{1}{c}{$M_P$} &
\multicolumn{1}{c}{$M_O$} &
\multicolumn{1}{c}{$N$} &
\multicolumn{1}{c}{$T_P$} &
\multicolumn{1}{c}{$T_O$} \\
\hline
     1    &      1    &     7.4076e-14 & -1.7347e-18 & -1.2872e+01 &  1.0685e-13 &  3.4528e+01 & -9.4104e-14 \\
     2    &      3    &    -5.6336e-01 &  1.6805e-17 &  4.8222e+00 & -3.0428e+00 & -2.9397e+01 & -1.2519e+00 \\
     3    &      5    &     5.4213e+00 &  1.0022e-15 &  3.4251e+00 & -9.2377e+00 &  6.5325e+00 &  1.2047e+01 \\
     4     &     7    &     7.3332e-01 & -1.4565e-15 & -1.8128e+01 &  4.8039e+01 & -1.7113e+01 &  1.6296e+00 \\
     5     &      9    &     5.4272e+00 & -4.1278e-15 & -9.9242e+00 & -6.5908e+00 &  1.3330e+00 &  1.2060e+01 \\
     6    &     11    &     1.0424e+01 &  0.0000e+00 & -7.6836e+00 &  6.3231e+00 &  3.9569e+00 & -1.1582e+01 \\
     7    &     13    &    -3.5769e+00 &  1.5543e-15 & -4.9466e+00 &  2.0936e+00 & -4.6705e+00 &  2.2284e+00 \\
     8    &     15    &    -3.9950e+00 & -8.8818e-16 & -2.2732e+01 &  2.6003e+01 & -1.7988e+01 & -5.7561e+00 \\
     9    &     17    &     2.8175e+00 & -8.8818e-16 & -1.1869e+01 & -5.1371e+00 &  2.7349e+01 &  9.4049e+00 \\
    10    &     19    &    -1.6165e+00 & -2.2204e-16 & -4.4764e+00 &  2.7492e+00 & -1.1102e-16 &  1.7961e+00 \\
    11    &     21    &     1.1450e+01 &  6.2172e-15 & -9.5570e+00 &  2.5956e+01 &  0.0000e+00 & -9.3528e+00 \\
    12    &     23    &     5.0135e+00 & -3.5527e-15 & -4.8249e+01 & -4.9428e+01 &  1.7764e-15 &  9.4320e-01 \\
    13    &     25    &    -1.9465e+00 & -1.7764e-15 & -7.6578e+00 & -4.4980e+00 & -2.2204e-16 &  2.1628e+00 \\
    14    &     27    &    -6.7125e+00 &  7.1054e-15 & -2.7835e+01 & -2.8757e+01 &  1.7764e-15 &  1.0843e+01 \\
    15     &    29    &    -5.0427e-01 & -3.6499e-15 & -1.8494e+01 & -1.8938e+01 &  1.7347e-17 &  1.4389e+00 \\
    16    &      2    &    -5.8588e+00 &  0.0000e+00 & -9.6630e+00 & -6.6011e+00 &  1.2360e+00 & -1.2028e+01 \\
    17    &      3    &    -5.4813e+00 &  1.9984e-15 & -2.7681e+00 & -4.1040e+00 &  4.4409e-16 & -1.3757e+01 \\
    18    &      4    &     2.4904e+00 & -4.4409e-16 &  3.9431e+00 & -1.0157e+00 & -1.1102e-16 &  5.1129e+00 \\
    19    &      5    &    -4.2331e+00 & -4.4409e-16 &  4.1806e+00 &  1.1984e+01 &  8.8818e-16 &  1.5404e+01 \\
    20    &      6    &     6.6230e+00 & -5.3291e-15 & -3.2485e+00 &  5.1667e+00 &  0.0000e+00 &  1.3598e+01 \\
    21    &      7    &     8.6430e-01 &  2.7200e-15 & -2.1780e+01 & -4.9983e+01 &  3.5527e-15 &  5.0337e+00 \\
    22    &      8    &    -1.7673e+01 &  4.4409e-15 & -1.1856e+01 &  3.1501e+01 &  1.7764e-15 & -3.6284e+01 \\
    23    &      9    &    -1.1982e+01 & -8.8818e-16 & -1.1818e+01 &  1.0996e+01 &  2.7234e+01 & -4.7865e-01 \\
    24    &     12    &    -5.4276e+00 & -8.8818e-16 & -1.9876e+00 &  2.7980e+00 & -1.0520e+00 & -1.3596e+00 \\
    25    &     13    &     1.1686e+00 & -7.7716e-16 & -4.8261e+00 & -7.2353e+00 & -4.4409e-16 &  6.5330e-01 \\
    26    &     14    &     5.4303e+00 & -6.4393e-15 & -2.1667e+00 & -4.4747e+00 &  8.8818e-16  &-1.8479e+01 \\
    27    &     15    &    -3.2319e+00 & -8.8818e-15 & -1.0783e+01 & -3.7577e+01 &  3.5527e-15  &-1.6830e+01 \\
    28    &     16    &    -3.1680e+01 &  3.5527e-15 & -1.5370e+01 & -1.8526e+01 & -1.3323e-15  & 1.6389e+01 \\
    29    &     17    &    -6.4497e+00 &  1.7764e-15 & -2.6942e+01 &  4.8428e+01 & -9.9149e+00  &-7.0609e-01 \\
    30    &     20    &    -5.4470e+00 &  8.8818e-16 & -1.7826e+00 & -9.9605e-02 &  1.7214e+00  & 4.6837e+00 \\
    31    &     21    &     4.3794e-02 &  8.8818e-16 & -1.3275e+01 & -1.4048e+01 & -4.4409e-16  & 3.4600e+00 \\
    32    &     22    &    -2.1003e+01 &  5.3291e-15 & -6.7778e+00 &  3.3121e+00 & -1.7764e-15  & 1.4138e+01 \\
    33    &     23    &     1.3449e+00 &  7.1054e-15 & -1.5962e+01 &  2.9198e+01 & -2.4477e+00  & 9.2751e+00 \\
    34    &     26    &    -9.2472e+00 &  5.1070e-15 & -4.5523e+00 & -7.4792e+00 &  7.1545e+00  & 1.5605e+01 \\
    35    &     27    &    -1.0991e+00 & -9.7700e-15 & -1.0132e+01 &  1.8937e+01 & -3.3732e+00  & 1.1280e+00 \\
\hline\hline
\end{tabular}
\normalsize
\end{table}

\begin{table}
\caption{Reaction forces at supports, and panels which they are applied to,  for the origami structure in Figure\,\ref{yoshi_load}.}
\label{tab:yoshi:ext}
\centering
\small
\renewcommand{\arraystretch}{1.25}
\begin{tabular}{c c l l l }
\hline\hline
\multicolumn{1}{c}{vertex} &
\multicolumn{1}{c}{panel} &
\multicolumn{1}{c}{$F_x$} &
\multicolumn{1}{c}{$F_y$} &
\multicolumn{1}{c}{$F_z$} \\
\hline
                1  &    1 &  0.0000e+00 & -1.6952e-02 & -3.4528e+01 \\
                2  &    2 &  0.0000e+00 & -1.3689e+01 &  3.5919e+01 \\
                3  &    3 &  0.0000e+00 &  1.4258e+01 &  2.7823e+01 \\
                4  &    4 &  0.0000e+00 &  3.6695e+00 & -2.9009e+01 \\
                5  &    5 &  0.0000e+00 & -3.8246e+00 & -1.9461e+00 \\
                6  &    6 &  -4.0523e-14 & -7.7983e-01 &  4.5555e+00 \\
                7  &    7 &  0.0000e+00 & -3.8797e+00 & -2.0152e+00 \\
                8  &    8 &  0.0000e+00 &  3.7173e+00 & -2.9168e+01 \\
                9  &    9 &  0.0000e+00 &  1.4289e+01 &  2.8036e+01 \\
               10  &   10 &  0.0000e+00 & -1.3744e+01 &  3.5913e+01 \\
               10  &   11 &  0.0000e+00 &  3.1084e-04 & -3.4580e+01 \\
\hline\hline
\end{tabular}
\normalsize
\end{table}

\begin{table}
\caption{Vertex coordinates for the Kresling module in singular configuration in Figure~\ref{kresling3}\,(a).}
\label{tab:kres:ind:ref}
\centering
\small
\renewcommand{\arraystretch}{1.25}
\begin{tabular}{c l l l}
\hline\hline
\multicolumn{1}{c}{vertex} &
\multicolumn{1}{c}{$x$} &
\multicolumn{1}{c}{$y$} &
\multicolumn{1}{c}{$z$} \\
\hline
   1 &  1.0000e+01 &  0.0000e+00     &       0  \\
   2 & -5.0000e+00 &  8.6603e+00     &       0  \\
   3 & -5.0000e+00 & -8.6603e+00     &       0  \\
   4 & -8.6603e+00 &  5.0000e+00     &      20  \\
   5 & -0.0000e+00 & -1.0000e+01     &      20  \\
   6 &  8.6603e+00 &  5.0000e+00     &      20  \\
\hline\hline
\end{tabular}
\normalsize
\end{table}

\begin{table}
\caption{Internal actions in the self-stress state of the Kresling module in singular configuration shown in Figure\,\ref{kresling3}\,(a)). Only the values at two hinges are shown, the remaining hinges feature the same values according to the three-fold cyclic symmetry of the assembly.}
\label{tab:kres:ind:int}
\centering
\small
\renewcommand{\arraystretch}{1.25}
\begin{tabular}{c c l l l l l l}
\hline\hline
\multicolumn{1}{c}{hinge} &
\multicolumn{1}{c}{panel} &
\multicolumn{1}{c}{$M_T$} &
\multicolumn{1}{c}{$M_P$} &
\multicolumn{1}{c}{$M_O$} &
\multicolumn{1}{c}{$N$} &
\multicolumn{1}{c}{$T_P$} &
\multicolumn{1}{c}{$T_O$} \\
\hline
     1     &     1    &    -1.3517e-01 &  0.0000e+00 &  5.2110e-01 &  3.3945e-03 & -5.2234e-02 & -1.3086e-02 \\
     2     &     2    &     1.8194e-01 &  5.5511e-17 &  6.7783e-02 & -3.5127e-02 & -3.8808e-02 & -1.3086e-02 \\
\hline\hline
\end{tabular}
\normalsize
\end{table}

\begin{table}
\caption{Vertex coordinates for the Kresling module in Figure~\ref{kresling3}\,(b).}
\label{tab:kres:ref}
\centering
\small
\renewcommand{\arraystretch}{1.25}
\begin{tabular}{c l l l}
\hline\hline
\multicolumn{1}{c}{vertex} &
\multicolumn{1}{c}{$x$} &
\multicolumn{1}{c}{$y$} &
\multicolumn{1}{c}{$z$} \\
\hline
   1  & 1.0000e+01 &  0.0000e+00  &          0  \\
   2  &-5.0000e+00 &  8.6603e+00  &          0  \\
   3  &-5.0000e+00 & -8.6603e+00  &          0  \\
   4  &-7.6604e+00 &  6.4279e+00  &         20  \\
   5  &-1.7365e+00 & -9.8481e+00  &         20 \\ 
   6  & 9.3969e+00 &  3.4202e+00  &         20 \\
\hline\hline
\end{tabular}
\normalsize
\end{table}

\begin{table}
\caption{Internal actions for the Kresling module in Figure\,\ref{kresling3}\,(b). Only the values at two hinges are shown, the remaining hinges feature the same values according to the three-fold cyclic symmetry of the assembly.}
\label{tab:kres:int}
\centering
\small
\renewcommand{\arraystretch}{1.25}
\begin{tabular}{c c l l l l l l}
\hline\hline
\multicolumn{1}{c}{hinge} &
\multicolumn{1}{c}{panel} &
\multicolumn{1}{c}{$M_T$} &
\multicolumn{1}{c}{$M_P$} &
\multicolumn{1}{c}{$M_O$} &
\multicolumn{1}{c}{$N$} &
\multicolumn{1}{c}{$T_P$} &
\multicolumn{1}{c}{$T_O$} \\
\hline
     1    &      1    &    -1.8587e+00 & -2.2204e-16 &  1.0327e+01 &  6.5668e-02 & -1.1204e+00 & -1.8313e-01 \\
     2    &      2    &     4.7229e+00 & -4.4409e-16 & -8.5253e+00 & -1.4225e+00 & -1.5574e+00 & -3.4417e-01 \\
\hline\hline
\end{tabular}
\normalsize
\end{table}

\begin{table}
\caption{Reaction forces at supports, and panels which they are applied to, for the Kresling module in Figure\,\ref{kresling3}\,(b).}
\label{tab:kres:ext}
\centering
\small
\renewcommand{\arraystretch}{1.25}
\begin{tabular}{c c l l l }
\hline\hline
\multicolumn{1}{c}{vertex} &
\multicolumn{1}{c}{panel} &
\multicolumn{1}{c}{$F_x$} &
\multicolumn{1}{c}{$F_y$} &
\multicolumn{1}{c}{$F_z$} \\
\hline
                1    &  1 &  0.0000e+00 & 0.0000e+00 & 1.0000e+00 \\
                3    &  2 &  2.5530e-14 &  -0.0000e+00 & 1.0000e+00 \\
                5    &  3 &  7.5384e-15 &  -1.8445e-14 & 1.0000e+00 \\
\hline\hline
\end{tabular}
\normalsize
\end{table}

\begin{table}
\caption{Internal actions for the Kresling two-module assembly in Figure\,\ref{kresling3multi}\,(a). Only the values at five hinges are shown, the remaining hinges feature the same values according to the three-fold cyclic symmetry of the assembly.}
\label{tab:k2}
\centering
\small
\renewcommand{\arraystretch}{1.25}
\begin{tabular}{c c l l l l l l}
\hline\hline
\multicolumn{1}{c}{hinge} &
\multicolumn{1}{c}{panel} &
\multicolumn{1}{c}{$M_T$} &
\multicolumn{1}{c}{$M_P$} &
\multicolumn{1}{c}{$M_O$} &
\multicolumn{1}{c}{$N$} &
\multicolumn{1}{c}{$T_P$} &
\multicolumn{1}{c}{$T_O$} \\
\hline
     1     &     1    &     6.2104e+01 &  4.4409e-15 &  2.0815e+01 & -9.6760e-01 & -3.2260e+00 &  6.1189e+00 \\
     2     &     2    &    -7.5479e+01 & -7.1054e-15 & -4.1662e+01 &  9.9233e-01 & -3.8436e+00 & -6.1888e+00 \\
     7     &     2    &     1.3951e+00 & -3.1530e-14 &  8.5471e+00 & -8.6786e-01 &  4.4409e-16 & -1.1528e+01 \\
    10     &     7    &     7.0726e+01 & -6.2172e-15 &  1.9183e+00 &  1.1567e+00 & -2.2204e-16 & -6.6462e+00 \\
    11     &     8    &     7.8295e+01 &  2.8422e-14 &  5.2011e+01 &  4.5028e+00 &  2.2204e-16 & -5.3394e+00  \\
\hline\hline
\end{tabular}
\normalsize
\end{table}

\begin{table}
\caption{Internal actions for the Kresling three-module assembly in Figure\,\ref{kresling3multi}\,(b). Only the values at eight hinges are shown, the remaining hinges feature the same values according to the three-fold cyclic symmetry of the assembly.}
\label{tab:k3}
\centering
\small
\renewcommand{\arraystretch}{1.25}
\begin{tabular}{c c l l l l l l}
\hline\hline
\multicolumn{1}{c}{hinge} &
\multicolumn{1}{c}{panel} &
\multicolumn{1}{c}{$M_T$} &
\multicolumn{1}{c}{$M_P$} &
\multicolumn{1}{c}{$M_O$} &
\multicolumn{1}{c}{$N$} &
\multicolumn{1}{c}{$T_P$} &
\multicolumn{1}{c}{$T_O$} \\
\hline
     1    &      1    &    -6.7456e+01 & -9.3259e-15 & -7.4600e-01 &  1.1567e+00 &  6.6613e-16 & -6.6462e+00 \\
     2    &      2    &     8.2715e+01 &  1.4211e-14 &  3.3745e+01 & -4.5028e+00 & -6.6613e-16 &  5.3394e+00 \\
     7    &      2    &    -1.3951e+00 & -1.6431e-14 & -8.5472e+00 &  2.8418e+00 & -4.4409e-16 &  1.1206e+01 \\
    10    &      7    &    -4.4184e+00 &  2.0872e-14 &  1.2322e+01 & -4.3479e+00 &  2.3759e+00 &  1.2125e+01 \\
    11    &      8    &    -1.6417e+02 & -1.4211e-14 & -3.9722e+01 & -2.4649e+00 &  3.9435e+00 & -5.9641e-01 \\
    16    &      13    &     1.3945e+00 &  2.4425e-14 &  8.5472e+00 & -2.8418e+00 & -2.6645e-15 & -1.1206e+01 \\
    19    &     14    &     -6.7457e+01 & -3.9968e-15 & -7.4586e-01 &  1.1567e+00 &  4.4409e-16 & -6.6462e+00 \\
    20    &     15    &     -8.2715e+01 &  7.1054e-15 & -3.3745e+01  & 4.5028e+00  & 4.4409e-16 & -5.3394e+00 \\
\hline\hline
\end{tabular}
\normalsize
\end{table}

\end{document}